\documentclass[%
 amsmath,amssymb,
 pre,
twocolumn
]{revtex4-2}

\usepackage{graphicx}% Include figure files
\usepackage{dcolumn}% Align table columns on decimal point
\usepackage{bm}% bold math
\usepackage{hyperref}% add hypertext capabilities
\usepackage{pdfcomment}
\usepackage{svg}
\usepackage{float}
\usepackage{amsthm}
\usepackage{soul}
\usepackage{xcolor}
\usepackage{colortbl}

\usepackage{tabularx}
\usepackage{booktabs}
\usepackage{multirow}
\usepackage{array}
\usepackage{makecell}

\usepackage[colorinlistoftodos]{todonotes}

\newcolumntype{C}{>{\centering\arraybackslash}X}

\begin{document}

\title{Heterophily as a generative mechanism for self-organized synergistic interdependencies}

\author{Enrico Caprioglio}
 \email{Contact author: e.caprioglio@sussex.ac.uk}
 \affiliation{%
  Department of Informatics, University of Sussex, Brighton, United Kingdom
  }%
\author{Luc Berthouze}%
  \affiliation{%
  Department of Informatics, University of Sussex, Brighton, United Kingdom
  }%

\date{\today}

% abstract
\begin{abstract}
Understanding what and how causal dynamical mechanisms generate collective phenomena is a central challenge in complexity science.
Recent studies have focused on identifying the mechanisms underlying the synergistic interdependencies that characterise these phenomena in systems with fixed interaction structures.
Yet, real-world systems displaying collective phenomena, such as brains, societies, and ecosystems, are adaptive: interactions change in time.
Here, we show that heterophily is a minimal local adaptive mechanism for the emergence of self-organized synergistic interdependencies.
We study a paradigmatic spin-glass-like model with co-evolving couplings to show how heterophily generates the conditions for synergy to emerge.
By solving the minimal $N=3$ case analytically, we reveal the precise mechanism: heterophily weakens pairwise dependencies while inducing high-order dependencies via geometric constraints on the configurations it selects.
Together, these two effects underpin synergy.
Numerical simulations confirm that this mechanism persists in large systems and that it is robust under parameter heterogeneities and dynamics.
We demonstrate the applicability of our results by showing how heterophily can disrupt polarization while promoting synergistic information dynamics of opinions, where individuals' opinions are better explained by group-level influences than by pairwise ones.
These results offer a parsimonious route to self-organized synergistic interdependencies in information-processing systems, with potential applications in computational social science, neuroscience, and biology.
\end{abstract}

\maketitle

% intro
\section{Introduction}\label{paper3:sec:introduction}
At its core, complexity science is the systematic study of when and how collective phenomena arise~\cite{Jensen2023Emergence}.
These phenomena often manifest as high-order non-local interdependencies that are irreducible to low-order relationships~\cite{battiston2022high-order-book}.
Although such interdependencies are increasingly observed across brains~\cite{Borsboom2013psychopatology, Sporns2022brainIT, Barabsi2023needNetwork}, ecosystems~\cite{Sol2022ecologyNext} and societies~\cite{Caldarelli2023digitalTwins}, their mechanistic origins remain poorly understood.
A necessary step to solve this problem is to identify the mechanisms that generate these non-local interdependencies in simple, analytically tractable models of complex networked systems~\cite{Rosas2022disentangling,Malizia2024reconstructingHighOrder,Robiglio2025synergistic}.

Advances in multivariate Information Theory (IT) provide precise mathematical tools to quantify statistical interdependencies irreducible to pairwise statistics~\cite{Williams2010pid,Rosas2019oInfo,Rosas2025entropic}.
Synergy quantifies information that is available only from the joint observation of multiple variables (non-local).
In contrast, redundancy is information that can be disclosed by any subset of such variables (local and overlapping).
Intuitively, synergy is associated with weak low-order dependencies coexisting with strong high-order ones (e.g., $I(X_1;X_2)\approx0$ and $I(X_1;X_2|X_3)>0$ where $I(\cdot\,;\,\cdot)$ denotes the mutual information), whereas redundancy reflects strong low-order dependencies~\cite{Rosas2025entropic}.
Prior work has identified two simple causal mechanisms for the occurrence of synergies: explicit high-order interactions~\cite{Rosas2019oInfo,Robiglio2025synergistic}, and frustrated interaction patterns in systems without high-order interactions~\cite{Caprioglio2026synergisticMotifs}.
Crucially, these mechanisms show when synergy can arise given a prescribed (time-independent) interaction structure, but not how synergies emerge in a self-organized manner.

Adaptivity is a foundational concept in complexity science~\cite{holland1962adaptive,Gross2007adaptive}: interactions between elements change as a function of the elements' activity.
Notably, synergy has been reported in several systems in which interactions are adaptive.
For example, in brains it has been associated with evolution~\cite{Luppi2022synergisticCore}, development~\cite{Varley2025HighOrderBrainDevelopment} and ageing~\cite{Gatica2021HighOrderAging}, and has also been shown to be modulated by transcranial ultrasound stimulation~\cite{Gatica2025plasticitySynergy, Gatica2024tusPlasticity}.
In artificial systems, synergy only appears after learning in LLMs~\cite{urbina2026brain} and increases in artificial neural networks as they learn to solve multiple multimodal/integrative tasks~\cite{Proca2024synergyNeuralNetworks}.
While these observations do not imply a single universal adaptive mechanism underlying the emergence of synergy, they highlight a current crucial gap in the literature:
a theory of how adaptive interaction mechanisms can support or give rise to synergistic interdependencies is missing.
As a first step, here we ask what local adaptive dynamical rules can lead to the emergence of synergy.

Consider the most widely studied local adaptive rule across fields: homophily~\cite{McPherson2001homophily} (``like attracts like'' or ``cells that fire together wire together'').
In general, homophily takes an attraction-repulsion form: links are reinforced when similarity between elements is high, and weakened when dissimilarity is high~\cite{FLACHE2011attractionRepulsion, Bail2018attractionRepulsion}.
Intuitively, this amplifies low-order (pairwise) dependencies, which is a signature of redundancy rather than synergy.
Additionally, homophilous dynamical rules can lead to the emergence of balanced macroscopic interaction patterns~\cite{Pham2022homophily, Korbel2023selfAssembly}.
These self-organized balanced patterns impose triadic constraints which align pairwise relations across multiple variables, thereby increasing information overlap~\cite{Caprioglio2026synergisticMotifs}.
Because of this, we expect that within this class of local, pairwise adaptive rules, homophily promotes locally accessible redundant interdependencies, rather than non-local, synergistic interdependencies.

In contrast to balanced systems, frustrated ones tend to give rise to synergistic interdependencies~\cite{Matsuda2000pairwise, Rosas2019oInfo, Rosas2025entropic,Caprioglio2026synergisticMotifs}.
Frustrated interaction patterns prevent the system's elements from satisfying local (pairwise) constraints simultaneously, thereby suppressing low-order dependencies.
Yet, they preserve high-order dependencies across groups of variables due to the imposed (fixed) interaction patterns which enforce frustration.
This combination, weak low-order dependencies coexisting with high-order ones, is precisely the hallmark of synergy.
% (see also Appendix ? from the previous chapter).
However, frustration is typically studied as a property of systems with prescribed interaction patterns.
Can such frustration-like high-order dependencies arise from simple, local adaptive rules and give rise to synergy?
Heterophily, the tendency to reinforce ties with dissimilar elements and weaken those with similar ones~\cite{Motsch2014heterophily}, is a natural candidate:
its competing attractive and repulsive tendencies naturally suppress pairwise dependencies, yet the problem of satisfying them all across pairs is geometrically constrained.
In this work, we hypothesise that heterophily is a sufficient minimal pairwise adaptive mechanism for the emergence of self-organised synergistic interdependencies.

Spin glasses are paradigmatic models of complex systems~\cite{Stein2013spinGlasses}.
Using an extension of these models with co-evolving adaptive couplings $J_{ij}(t)$~\cite{Pham2022homophily, Korbel2023selfAssembly, Thurner2025polarization}, we show precisely how heterophily creates the conditions for synergy by weakening pairwise dependencies while giving rise to genuine high-order structure not captured by pairwise statistics.
We solve the model exactly for small systems to isolate and identify the underlying fundamental mechanisms, and extend these insights numerically to larger systems.

Such abstract models are not designed for quantitative fits to real-world systems, yet they provide important mechanistic insight into the principles that may generate complex behaviour in real-world systems~\cite{Galesic2025experimentHomophily,Thurner2025polarization}.
To demonstrate the insight that our abstract model can provide, we verify the robustness of our analytical insights under alternative update schemes (which can break detailed balance) and heterogeneous parameters commonly used in computational social science.
In these settings, we show that heterophily can disrupt polarization and promote synergistic information dynamics, in which opinion formation is influenced more strongly by group-level mechanisms than by pairwise ones.
Finally, we discuss how these results may inform a broad class of canonical (discrete) models with applications spanning neuroscience, computational social science, and evolutionary biology.
% methods
\section{Model and measures}\label{paper3:sec:methods}
\subsection{Homophily vs heterophily}
\begin{figure}
    \centering
    \includegraphics[width=.99\linewidth]{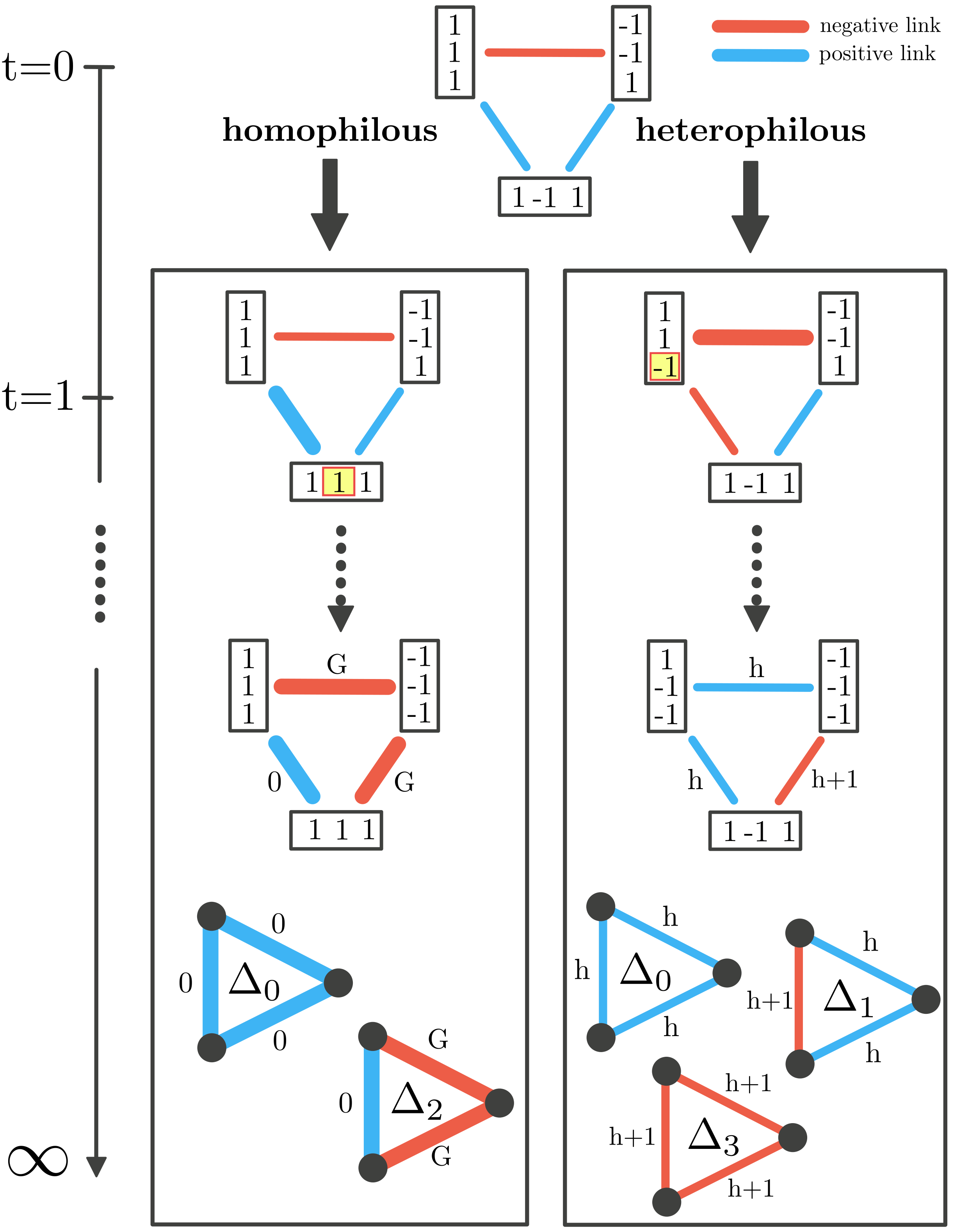}
    \caption[Homophily strengthens pairwise interactions while heterophily weakens pairwise interactions]{\footnotesize Homophily strengthens pairwise interactions while heterophily weakens pairwise interactions.
    Schematic depiction of typical homophilous (\textbf{left}) and heterophilous (\textbf{right}) dynamics.
    Both dynamics start from the same initial condition ($t=0$) with $N=3$ nodes, each associated with a $G$-dimensional spin vector ($G=3$ such that $h=1$).
    Blue (red) edges denote positive (negative) adaptive links, where the edge thickness is proportional to the overlap $|O_{ij}|$.
    At each step (e.g., at $t=1$ where the node highlighted in red is updated), the homophilous dynamics tends to increase similarity and dissimilarity ($|O_{ij}|=|G|$), while heterophily tends to minimize them ($|O_{ij}|=h\bigvee{h+1}$).
    At $t=\infty$, we display the most likely, metastable, configurations (corresponding to the zero-temperature ground states) which depend on the value of $\alpha$ (not shown).
    At $t=\infty$, we omit the node states to highlight the patterns of relationships between nodes that characterize these states (i.e., the high-order constraints).
    The label $\Delta_k$ denotes a triangle with $k$ negative interactions.
    If $k$ is even (odd) the triplet is said to be balanced (antibalanced).}
    \label{paper3:fig:intro_big_picture}
\end{figure}

% distances and similarity paragraph
To model homophily and heterophily from first principles, all we need is a measure of distance.
Consider a system of $N$ elements with underlying network of interactions described by the adjacency matrix $A$.
This matrix is static (non-adaptive), serving as the underlying substrate for all possible connections between elements.
Each element is associated with a $G$-dimensional spin $\mathbf{s}_i(t)\in\{\pm1\}^G$.
Then, the degree of alignment between elements is given by the dot product
\[
O_{ij}:=\mathbf{s}_i(t)\cdot \mathbf{s}_j(t)= G-2d_{ij},
\]
where $d_{ij}:=d(i,j)$ is the Hamming distance between $\mathbf{s}_i$ and $\mathbf{s}_j$ (by remapping $-1\to{0}$).
In this minimal setting, two elements are said to be ``similar'' if $O_{ij}>0$ and ``dissimilar'' if $O_{ij}<0$. Defining $h=(G-1)/2$, similarity corresponds to $d_{ij}\leq{h}$, dissimilarity to $d_{ij}\geq{h+1}$. As in~\cite{Pham2022homophily}, we restrict our analysis to odd $G$, where $O_{ij}\in\{-G, -G+2,\,\dots,\,G\}\setminus\{0\}$ such that ``similar'' and ``dissimilar'' are unambiguous.

% adaptive couplings paragraph and local energy
Let $J_{ij}(t)=A_{ij}\,\mathrm{sign}(O_{ij}(t))$ define the nature of the adaptive coupling between elements $i$ and $j$ at time $t$.
Based on the model introduced in~\cite{Pham2022homophily}, the simplest model that allows us to implement all the attraction/repulsion tendencies can be written using the following local energy:
\begin{equation}
    E^{(i)} = \lambda_i\left(\frac\alpha{G}\sum_{j:J_{ij}>0}O_{ij} + \frac{1-\alpha}{G}\sum_{j:J_{ij}<0} |O_{ij}| \right), \label{paper3:eqn:local_hamiltonian}
\end{equation}
where $\alpha\in[0,1]$ controls the relative pressure exerted by similar vs dissimilar neighbours of $i$ ($\alpha=1$ ignores dissimilar neighbours, $\alpha=0$ ignores similar neighbours).
The parameter $\lambda_i\in\{\pm1\}$ toggles homophily vs heterophily:
\begin{itemize}
    \item for $\lambda_i=-1$ (homophily) we recover the model studied in~\cite{Pham2022homophily, Galesic2025experimentHomophily,Thurner2025polarization}, where decreasing $E^{(i)}$ increases alignment with similar neighbours (attraction to the similar) and decreases alignment with dissimilar ones (repulsion of the dissimilar).
    \item for $\lambda_i=1$ (heterophily) decreasing $E^{(i)}$ decreases alignment with similar neighbours (repulsion of the similar) and increases alignment with dissimilar ones (attraction to the dissimilar).
\end{itemize}
Intuitively, homophily promotes cohesion or division, depending on $\alpha$, by pushing distances between nodes to the extremes $d\in\{0,G\}$ (maximising $|O_{ij}|$, see Fig.~\ref{paper3:fig:intro_big_picture} left).
Heterophily instead promotes heterogeneity by pulling distances to the threshold $d\in\{h,h+1\}$ (minimising $|O_{ij}|$, see Fig.~\ref{paper3:fig:intro_big_picture} right).

% dynamics paragraph
Dynamics is defined by single-site Glauber updates for the global energy $E(S)=\tfrac12\sum_i E^{(i)}(S)$, where $S=\{s_i\}_{i=1}^N$ denotes the state of the whole system.
At each time step, a random node $i$ is selected and a random site in $\mathbf{s}_i$ is flipped.
Let $S'$ denote the proposed new state, with energy $E(S')$.
The proposed update is then accepted with probability
\begin{equation}
    p(S \to S') = \frac{1}{1+\exp(\beta \Delta E)},
    \label{paper3:eqn:transition-probability}
\end{equation}
where $\Delta E = E(S') - E(S)$ and $\beta$ is the inverse temperature.
Because the proposal is symmetric, this dynamics satisfies detailed balance with respect to the Boltzmann distribution $\propto e^{-\beta E}$.
This allows us to study the model analytically for small $N$.
In Sec.~\ref{paper3:sec:robustness_analysis}, to test robustness, we also consider a local-update variant based on the local energy $E^{(i)}$.
In Sec.~\ref{paper3:sec:disrupting_polarization}, for direct comparability with Refs.~\cite{Pham2022homophily,Thurner2025polarization}, we instead use the Metropolis acceptance rule with local energy changes.

% Dynamics is defined by single-site Metropolis updates for the global energy $E(S)=\tfrac12\sum_iE^{(i)}(S)$, where $S=\{s_i\}_{i=1}^N$ denotes the state of the whole system.
% At each time step, a random node $i$ is selected and a random site in $\mathbf{s}_i$ is flipped such that $\mathbf{s}_i\to\tilde{\mathbf{s}}_i$. Let the new proposed state of the system be $S'$ with energy $E(S')$.
% Then, the proposed state is accepted with probability
% \begin{equation}
%     p(S\to S') = \mathrm{min}\left\{1, \exp(-\beta\Delta{E})\right\}
%     \label{paper3:eqn:transition-probability}
% \end{equation}
% where $\Delta{E}=E(S') - E(S)$ and $\beta$ is the inverse temperature.
% Because the proposal is symmetric, this dynamics satisfies detailed balance with respect to the Boltzmann distribution $\propto e^{-\beta E}$.
% This allows us to study the model analytically for small $N$.

\subsection{Information-theoretic measures}\label{paper3:sec:IT-preliminaries}
\subsubsection{O-information}
The O-information ($\Omega$) is a symmetric, signed multivariate extension of the mutual information that quantifies the relative dominance of synergistic vs redundant dependencies~\cite{Rosas2019oInfo}.
It can be written as
\begin{equation}
    \Omega(\mathbf{X}^N) = (N-2)H(\mathbf{X}^N) + \sum_{j=1}^N\left[H(X_j) - H(\mathbf{X}_{-j})\right],\label{paper3:eqn:O-info-full}
\end{equation}
where $H(\cdot)$ denotes the Shannon entropy, $\mathbf{X}^N=(X_1,\,\dots,\,X_N)$ denotes the whole system (a discrete-time stochastic process of $N$ variables) and $\mathbf{X}^{N-1}_{-j}=\mathbf{X}^N\setminus X_j$ denotes the system with the element $j$ removed.
When $\Omega>0$, the system's interdependencies are redundancy-dominated.
When $\Omega<0$ instead, the system's interdependencies are synergy-dominated.
Importantly, values of $\Omega$ near zero do not necessarily indicate the absence of high-order interdependencies, they can also indicate that redundant and synergistic interdependencies are equivalent in magnitude.

To see this more clearly, we follow the entropic-conjugation formalism of~\cite{Rosas2025entropic}. Measures of multivariate interdependencies based on the Shannon information (such as Eq.~\eqref{paper3:eqn:O-info-full}) can be rewritten in terms of the common (non-negative) basis $\{u_k\}_{k=1}^{N-1}$, where $u_k$ is the average mutual information between two variables conditioned on $k-1$ others.
For triplets, which is our main focus, only the first two basis terms are needed to compute $\Omega$:
\begin{align}
u_1(\mathbf{X}^3) &= \frac{1}{3}\sum_{i<j}I(X_i;X_j), \label{paper3:eqn:u1-component} \\
u_2(\mathbf{X}^3) &= \frac{1}{3}\sum_{i<j<k}I(X_i;X_j|X_k) \label{paper3:eqn:u2-component} \\
\Omega(\mathbf{X}^3) &= u_1(\mathbf{X}^3) - u_2(\mathbf{X}^3). \label{paper3:eqn:omega-decomposed}
\end{align}
Intuitively, $u_1$ quantifies interdependences at the level of pairwise marginals, whereas $u_2$ quantifies interdependences revealed once a third variable is taken into account.
Note, $\Omega\approx 0$ can arise either because both low- and high-order effects are weak ($u_1\approx u_2\approx 0$) or because strong pairwise and triadic contributions cancel each other ($u_1\approx u_2 \gg 0$).

\subsubsection{Total dynamical O-information}
The O-information quantifies the balance between redundancy and synergy in the equal-time distribution of a set of variables.
To extend this notion to time-lagged (directed) statistical dependencies in multivariate time series, Stramaglia \emph{et al.}~\cite{Stramaglia2021dynamicalOinfo} introduced the dynamical O-information, defined as the variation of O-information induced by adding a target variable to a set of sources, while conditioning out the target's own history~\cite{Stramaglia2021dynamicalOinfo,Robiglio2025synergistic}.
%  to discount shared information due to common past (or common inputs)

Let $Y$ be a target variable with future sample $Y(t) = y(t+1)$.
Define its history as 
$Y_0(t) = (y(t),y(t-1),\,\dots,\,y(t-\tau+1))$, where $\tau$ is the temporal horizon.
The dynamical O-information from $\mathbf{X}^N$ sources (evaluated at time $t$) to $Y$ is
\begin{equation}
    d\Omega_N(Y;\mathbf{X}^N)
    = (1-N)\,I\left(Y;\mathbf{X}^N\,|\, Y_0\right)
    +\sum_{j=1}^{N} I\left(Y;\mathbf{X}^N_{-j}\,|\, Y_0\right),
\label{paper3:eqn:dOmega}
\end{equation}
where $I(\cdot\,;\,\cdot\,|\,\cdot)$ denotes conditional mutual information.
Positive values $d\Omega_N>0$ indicate that the information about $Y$ provided by the sources is predominantly redundant (overlapping across sources), whereas $d\Omega_N<0$ indicates predominantly synergistic information (only accessible from sources jointly)~\cite{Stramaglia2021dynamicalOinfo,Robiglio2025synergistic}.

To quantify dynamical high-order interdependencies within a group of $N$ variables without choosing a privileged target, Robiglio \emph{et al.}~\cite{Robiglio2025synergistic} define the total (or symmetrized) dynamical O-information as the sum over all choices of target:
\begin{equation}
    d\Omega^{\mathrm{tot}}_{N}(\mathbf{X}^N)
    = \sum_{j=1}^{N} d\Omega_{N-1}(X_j;\mathbf{X}^N_{-j}).
\label{paper3:eqn:TotDynamicalOinfo}
\end{equation}
This expression is also signed: $d\Omega^{\mathrm{tot}}_{N}>0$ (respectively $<0$) indicates redundancy-dominated (respectively synergy-dominated) dynamical information sharing within the group~\cite{Robiglio2025synergistic}.

\subsection{Simulations and observables}
For systems with $N>3$ at low temperatures, the dynamics can be strongly glassy, with long-lived metastable plateaus and long autocorrelation times.
Consequently, time averaging along a single trajectory to estimate observables at equilibrium is both computationally inefficient and potentially misleading.

To numerically estimate observables at equilibrium, we use an ensemble (replica) average, similar to the methods used in~\cite{Rajpal2025infoTheoryEcosystemsEvolution}.
For each parameter set (and each network realisation), we run $R$ replicas with independent random initial conditions and random number seeds.
After a burn-in period of $t_b$ sweeps, we record a configuration snapshot $S^{(r)}(t_b)$ from each replica $r$.
To compute the total dynamical O-information, we collect two configuration snapshots, at $S^{(r)}(t_b)$ and $S^{(r)}(t_b+1)$.
All the observables considered in this work, whether structural properties of $J$ or information-theoretic quantities, are computed from replica ensembles.
In other words, we report observables computed over the quasi-stationary ensemble of configurations after $t_b$ sweeps. 
Unless otherwise stated, we use $t_b=100$ throughout.
State probabilities are estimated using the frequency of occurrence of the states in the ensemble.
Note that the microstate space grows as $2^{NG}$.
Thus, for IT quantities we limit our numerical analysis to $G=3$ and use replica ensembles of size $R=10^4$ to estimate equal-time probabilities and $R=10^5$ to estimate dynamical information-theoretic quantities.
In the Appendix Sec.~\ref{paper3:sec:sensitivity-analyses}, we show that the estimated observables are stable upon increasing $t_b$ and $R$.

\subsection{Data and Code Availability}
All data and presented in this study are reproducible using the source code available at~\cite{Caprioglio2026repoSelfOrganized}.
% results
\section{Results}
Even though the model is defined by local, pairwise terms, it can generate self-organized high-order structure~\cite{Pham2022homophily}.
Here, we show that this is because the feasible combinations of pairwise relationships within a triplet of nodes are geometrically constrained: three spin vectors in $\{\pm1\}^G$ cannot realise any triple of Hamming distances. This realisability constraint couples the three edges $J_{ij}(t)$ in a triangle, acting as an effective triadic interaction, without the need for explicit high-order interactions.

Our central result is that whilst homophily and heterophily exploit the same mechanism for the emergence of high-order structure, the emergent informational architecture is fundamentally different.
Homophily pushes pairwise distances to the extremes ($d \in \{0,G\}$).
This produces balanced triadic motif structures together with strong pairwise dependencies, resulting in a redundancy-dominated organisation of information.
Heterophily instead pulls distances toward the similarity threshold ($d\in\{h,h+1\}$), weakening pairwise dependencies.
Nevertheless, heterophily selects a restricted set of microstate configurations that satisfy the geometric constraints, which can't be explained by locally available information alone, resulting in a synergistic high-order organisation.

We demonstrate these claims in two steps (Sec.~\ref{paper3:sec:triadic_constraints} and Sec.~\ref{paper3:sec:IT_analysis}, respectively).
($i$) We solve the minimal $N=3$ system exactly, showing how the preferred dyadic distances imposed by the local dynamical rule must also satisfy geometric realisability constraints.
This restricts triplet configurations to a particular set of degenerate ground states, characterized by particular triangle motifs that depend on the model's parameters ($\{\lambda_i\}_{i=1}^N,\,\alpha,\,G$).
($ii$) We then characterise these regimes through the lens of information theory.
This analysis reveals that homophily yields strong overlapping pairwise dependencies,
whereas heterophily produces weak pairwise dependencies coexisting with high-order ones.

Then (Sec.~\ref{paper3:sec:robustness_analysis}), we test the robustness of our results under local Metropolis update schemes and beyond the symmetric cases, i.e., for heterogeneous populations in which both homophilous and heterophilous elements are present, and when each element possesses a random value of $\alpha_i$.
Finally (Sec.~\ref{paper3:sec:disrupting_polarization}), we extend a recent line of work that uses the symmetric homophilous model to study opinion formation in social systems~\cite{Pham2022homophily,Korbel2023selfAssembly,Galesic2025experimentHomophily,Thurner2025polarization}.
We study the effect of introducing heterophilous individuals to a polarized society and investigate the formation of opinions through the lens of the total dynamical O-information.

\subsection{Emergence of triadic constraints and macroscopic organisation}\label{paper3:sec:triadic_constraints}
\begin{figure}
    \centering
    \includegraphics[width=1\linewidth]{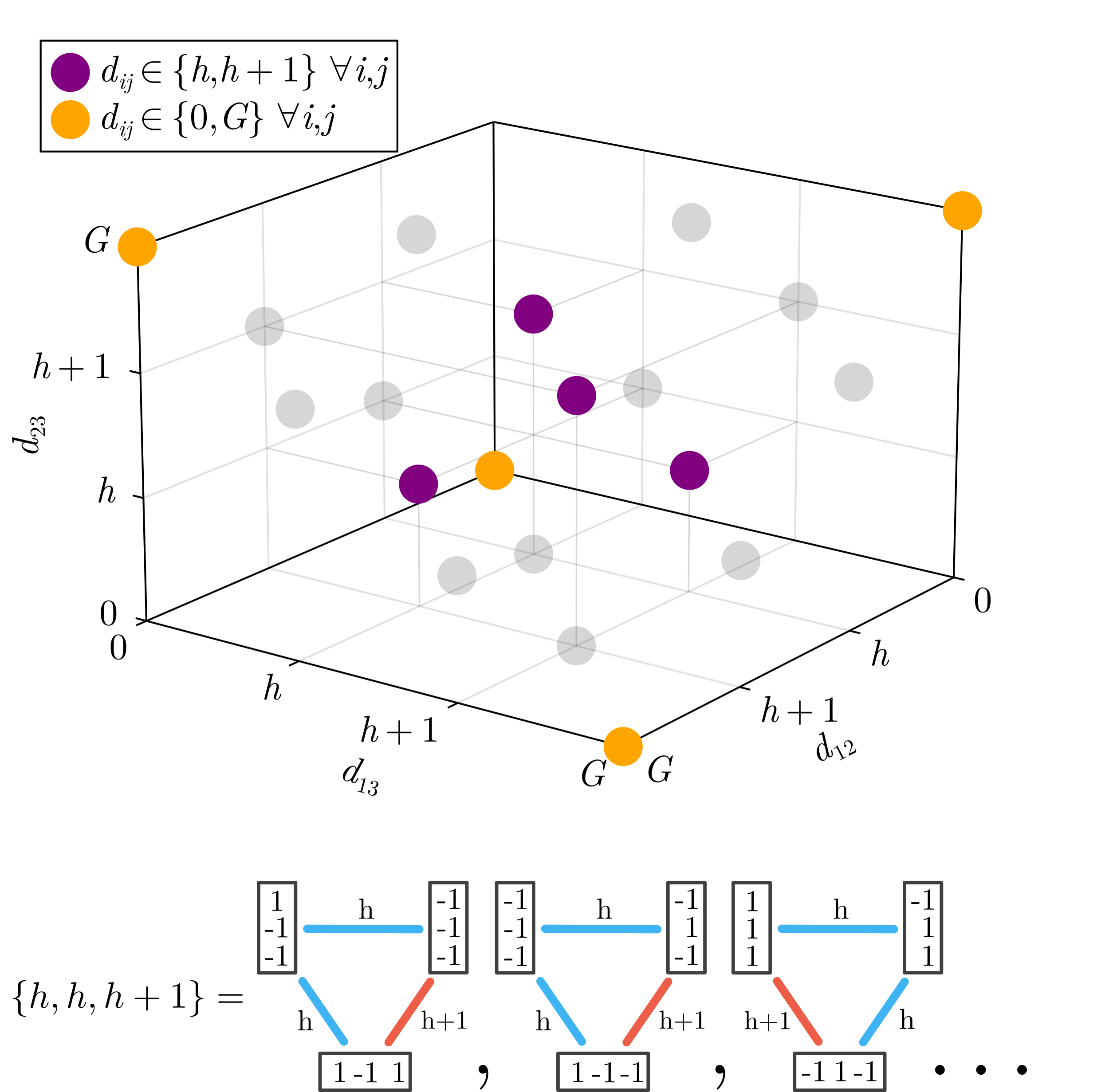}
    \caption[Distance state space: Homophily pushes distances to the extremes, heterophily pulls distances to the threshold]{Homophily pushes distances to the extremes, heterophily pulls distances to the threshold.
    \textbf{Top}: distance state space $\{d_{12},d_{13},d_{23}\}$ for $G=3$ and $N=3$.
    States that only have distances which maximise $|O_{ij}|$ are highlighted in orange.
    States that only have distances which minimize $|O_{ij}|$ are highlighted in purple.
    Grey markers denote all other possible states.
    \textbf{Bottom}: examples of microstates that are characterized by the same unordered triple of distances $\{h,h,h+1\}.$}
    \label{paper3:fig:spins_state_space_vs_distance}
\end{figure}

\begin{table*}[t]
\centering
\setlength{\tabcolsep}{6pt} % adjust spacing if needed
\renewcommand{\arraystretch}{1.15}

\begin{tabular}{@{} c c c c c @{}}
\toprule
$\boldsymbol{\lambda}$ pattern & $\alpha$ & $\delta^\star$ & $\Delta_k$ & $E$ \\
\midrule

\multirow{2}{*}{$(-,-,-)$}
& $\alpha\leq 1/2$ & $\{0,G,G\}$ & $\Delta_2$ & $-4+2\alpha$ \\
& $\alpha\geq 1/2$ & $\{0,0,0\}$ & $\Delta_0$ & $-6\alpha$ \\
\midrule

\multirow{3}{*}{\makecell[c]{$(+,+,+)$\\ with $h$ odd}}
& $\alpha\leq 1/4$ & $\{h-1,h,h\}$ & $\Delta_0$ & $\tfrac{10\alpha}{G}$ \\
& $1/4\leq\alpha\leq 1/2$ & $\{h,h,h+1\}$ & $\Delta_1$ & $\tfrac{2}{G}(\alpha+1)$ \\
& $\alpha\geq 1/2$ & $\{h+1,h+1,h+1\}$ & $\Delta_3$ & $\tfrac{6}{G}(1-\alpha)$ \\
\midrule

\multirow{3}{*}{\makecell[c]{$(+,+,+)$\\ with $h$ even}}
& $\alpha\leq 1/2$ & $\{h,h,h\}$ & $\Delta_0$ & $\tfrac{6\alpha}{G}$ \\
& $1/2\leq\alpha\leq 3/4$ & $\{h,h+1,h+1\}$ & $\Delta_2$ & $\tfrac{2}{G}(2-\alpha)$ \\
& $\alpha\geq 3/4$ & $\{h+1,h+1,h+2\}$ & $\Delta_3$ & $\tfrac{10}{G}(1-\alpha)$ \\
\midrule
\end{tabular}

\caption{For each symmetric $\lambda$ pattern for systems of size $N=3$, we report the ground states $\delta^\star$, written in the distance state space notation $\{d_{12},d_{13},d_{23}\}$, its associated SBT triangle characterization $\Delta_k$, where $k$ denotes the number of negative edges, and the ground state energy $E$.
For distinct values of $\alpha$ we have different regimes in which a particular configuration $\delta^\star$ is preferred (lowest energy).
}
\label{paper3:tab:groundstatesN3}
\end{table*}

\subsubsection{Ground state solutions for small systems}\label{paper3:sec:ground-states-small}
Here we show precisely how the preferred dyadic distances imposed by the local rule (homophily or heterophily) must also satisfy geometric realisability constraints, leading to particular sets of ground states that minimize $E(S)$.
For the sake of clarity, we report in the main text the results for symmetric systems, where all nodes are either heterophilous ($\lambda_i=1$ for all $i$) or homophilous ($\lambda_i=-1$ for all $i$).

For $N=3$ it is convenient to describe a microstate $S=(\mathbf{s}_1,\mathbf{s}_2,\mathbf{s}_3)\in\{\pm1\}^{N\times G}$ by the three pairwise distances
\[
\delta := \{d_{12},d_{13},d_{23}\},
\]
rather than by the individual spin components. Here $\delta$ is an unordered multiset, such that $\{a,b,c\}=\{b,a,c\}$ where $a,b,c\in\{0,1,\dots,G\}$ can take the same value. In other words, any permutation of $(d_{ij},d_{ik},d_{jk})$, with $i\neq{j}\neq{k}$ represents the same $\delta$.
Because the model is invariant under relabelling of nodes in the symmetric cases, $\delta$ uniquely identifies a particular energy level (up to microstate degeneracy).
In Fig.~\ref{paper3:fig:spins_state_space_vs_distance} (bottom) we show three examples of microstates with the same energy that can all be represented by the triplet of distances $\{h,h,h+1\}$.

Importantly, not every triple $\{d_{12},d_{13},d_{23}\}$ is realisable by three vertices of the $G$-dimensional hypercube.
A realisable triple must satisfy the hypercube constraints
\begin{eqnarray}
    d_{12}+d_{13}+d_{23} &\equiv& 0 \pmod 2, \label{paper3:eqn:parity}\\
    d_{12}+d_{13}+d_{23} &\le& 2G, \label{paper3:eqn:sum}\\
    d_{ij} \le d_{ik}+d_{jk}&\qquad& \text{for all distinct } i,j,k. \label{paper3:eqn:triangle}
\end{eqnarray}
In Fig.~\ref{paper3:fig:spins_state_space_vs_distance} (top), we visualize all the possible states (in terms of $\{d_{12}, d_{13}, d_{23}\}$) that satisfy these constraints for $G=3$.
We refer to the set of realisable distance triples $\delta$ satisfying Eqs.~\ref{paper3:eqn:parity}--\ref{paper3:eqn:triangle} as distance state space.

For the symmetric cases in which $\lambda_i=\lambda$ for all nodes $i$, the global energy can be more conveniently written as
\begin{equation}
    E(\delta)= \frac{2\lambda}{G}\sum_{i<j}f_{\alpha,G}\left(d_{ij}\right)
    \label{paper3:eqn:global_hamiltonian}
\end{equation}
where
\[
    f_{\alpha,G}(d) = \begin{cases}
    \alpha(G-2d) & \text{if }d\leq{h}, \\
    (1-\alpha)(|G-2d|) & \text{if }d\geq{h+1}.
    \end{cases}
\]
To find the ground states (i.e., triplet $\delta^\star$ in the distance state space that minimizes Eq.~\ref{paper3:eqn:global_hamiltonian}) we proceed in two steps.
First, given ($\lambda,\,\alpha, G$), we identify the distance $d$ that minimizes $\lambda f_{\alpha,G}(d)$.
Second, we check if the distance triplet $\delta$ in which each distance minimizes $f_{\alpha,G}(d)$ is allowed by imposing the geometric constraints (Equations~\eqref{paper3:eqn:parity}--\eqref{paper3:eqn:triangle}).
If the unconstrained triple is not allowed, we apply the smallest correction (i.e., we change one of the distances such that $\lambda (f_{\alpha,G}(d_{\mathrm{new}}) - f_{\alpha,G}(d_{\mathrm{old}}))$ is minimized).
This reduces the search to a few candidate triplets.
We further classify these triplets into balanced and antibalanced triangles using the structural balance theory standard notation $\Delta_k(t)$~\cite{Marvel2009energySBT,Pham2022homophily}, where $\Delta_k(t)$ denotes a triangle $J(t)\in\{\pm1\}^{3\times3}$ with $k$ negative couplings, i.e., the number of distances $d_{ij}\geq{h+1}$ (see for example Fig.~\ref{paper3:fig:intro_big_picture}).
Then, a triangle is balanced if $k$ is even and antibalanced otherwise.

In Table~\ref{paper3:tab:groundstatesN3} we report the ground states for the symmetric homophilous and heterophilous models. For completeness, we include the solutions for the heterogeneous triplets $(\lambda_1,\lambda_2,\lambda_3)$ in the Appendix A Sec.~\ref{paper3:sec:heterogeneous-lambda}.
For three homophilous nodes $(-,-,-)$, the ground states are always balanced:
depending on $\alpha$ it is either $\Delta_0$ with $\delta=\{0,0,0\}$, or $\Delta_2$ with $\delta=\{0,G,G\}$.
Notably, even when homophily locally pushes toward maximal dissimilarity, the geometric constraints force one edge to remain similar, thus resulting in a balanced motif.
For the symmetric heterophilous case instead $(+,+,+)$, the ground states can be either balanced or antibalanced depending on the value of $\alpha$.
In sum, these results demonstrate that particular triangle structures of interaction emerge not due to explicit high-order interactions, but due to the interplay between local preferred alignments and geometric constraints.

\subsubsection{Triangle analysis for large systems}\label{paper3:sec:triangle-analysis-large}
\begin{figure}
    \centering
    \includegraphics[width=1\linewidth]{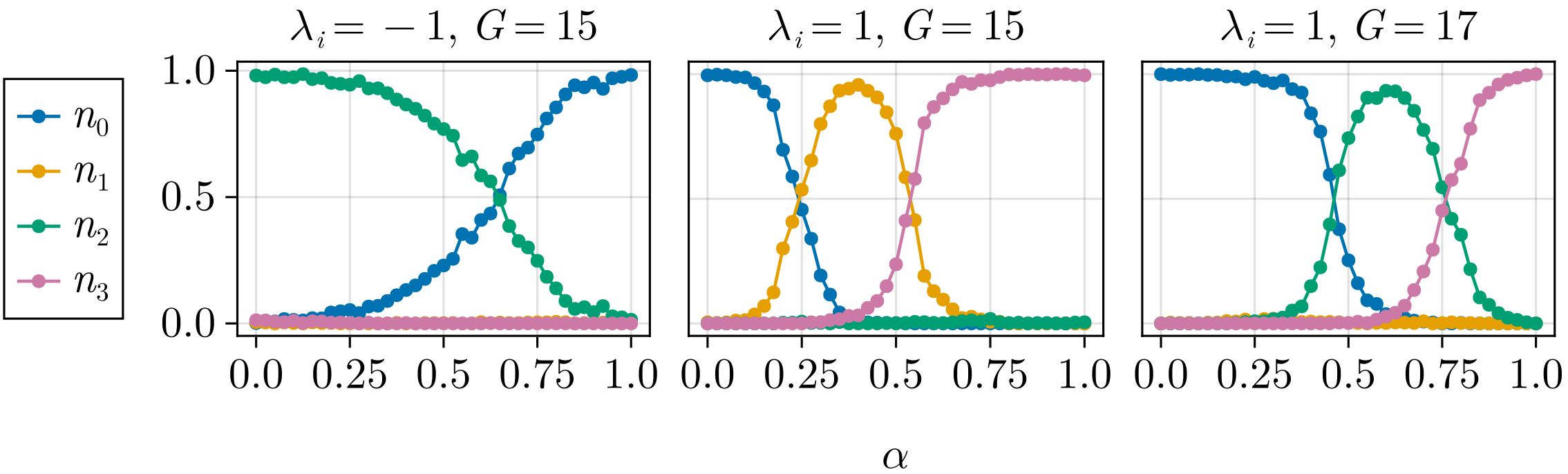}
    \caption[Mean triangle densities $n_k$ of triangle types $\Delta_k$ as a function of $\alpha$]{Mean triangle densities $n_k$ of triangle types $\Delta_k$ as a function of $\alpha$, where $k$ denotes the number of negative edges in a triangle.
    Results are averaged across $20$ independent realizations of small-world graphs with $N=50$, average degree $k_{\mathrm{avg}}=5$, and probability of rewiring $\epsilon=0.2$.
    Homophily ($\lambda_i=-1$ for all $i$, $\beta=15$) promotes balanced motifs, while heterophily ($\lambda_i=1$ for all $i$, $\beta=50$) can promote either balanced or antibalanced motifs depending on $\alpha$.}
    \label{paper3:fig:triangles_large_N_example}
\end{figure}
We now test whether large systems self-organize into structures in which particular triangles $\Delta_k$ are overrepresented. 
Intuitively, when a node is selected and a site $g$ in $\{s_i^g\}_{g=1}^G$ is flipped, the proposal is always accepted if it decreases the energy of each triangle that node $i$ is part of.
Thus, we expect this to lead to configurations in which the connected triangles satisfy the ground states we found for $N=3$.

Let $n_k(t)$ be the density of triangles $\Delta_k(t)$ with $k$ negative edges at some time $t$.
In Fig.~\ref{paper3:fig:triangles_large_N_example} we show $n_{k}$ as a function of $\alpha$ in systems of size $N=50$ with $G=15,\,17$ at low temperatures.
Each point is an ensemble average over $R=100$ independent replicas, where each measurement is taken at time $t=t_b$ after burn-in.

In the homophilous case ($\lambda_i=-1$ for all $i$) we recover the same results as shown in~\cite{Pham2022homophily}, where the majority of triangles are of the $\Delta_2$ kind for $\alpha<1/2$ and of the $\Delta_0$ kind for $\alpha>1/2$. In the heterophilous case ($\lambda_i=1$ for all $i$), we find that balanced triangles $\Delta_0$ dominate for small $\alpha$.
As $\alpha$ increases the system shifts toward motifs with more negative edges.
For intermediate $\alpha$, the parity constraints can force the system into either (balanced) $\Delta_2$- or (antibalanced) $\Delta_1$-dominated regimes depending on the parity of $h$, in great agreement with our $N=3$ ground state solutions.
For larger $\alpha$ instead, the system always results into an antibalanced $\Delta_3$-dominated system.

These results demonstrate that the ground state solutions for $N=3$ provide a powerful heuristic to predict the patterns of interaction emerging at large $N$.
Next we ask how these emerging structures relate to the informational architecture of the system.
\subsection{Information-theoretic analysis}\label{paper3:sec:IT_analysis}
Both homophily and heterophily generate non-trivial macroscopic interaction patterns.
However, these patterns alone do not determine whether the resulting informational architecture is organized redundantly or synergistically.
To distinguish these, we ask whether the observed triadic organization is already captured by pairwise mutual information, or whether it only emerges upon conditioning on the third variable, thereby revealing high-order dependence.
Specifically, we decompose triplet interdependencies into the low-order term $u_1$ (average pairwise mutual information) and the high-order term $u_2$ (average conditional mutual information), and study the O-information.

We study symmetric systems of size $N=3$ in two regimes.
First, we obtain an analytical expression for $u_1,\,u_2$, and $\Omega$ for the uniform distribution over the degenerate ground states at zero temperature ($T=0$, or $\beta\to\infty$).
Second, we compute these quantities exactly from the Boltzmann distribution at finite temperatures.
Since the ground-states have the largest Boltzmann weights at $T\neq0$, we expect the expressions at $T=0$ to provide a useful and interpretable approximation of the behaviours observed at finite non-zero $T$.
Finally, we confirm numerically that our analytical insights hold for large $N$ systems.

\subsubsection{High-order interdependencies in small systems}\label{paper3:sec:IT-small-systems}
Let $\delta^\star=\{a,b,c\}$ be a ground state solution for a given $(\lambda,\,\alpha,\,G)$ (see Table~\ref{paper3:tab:groundstatesN3}), and let
\begin{equation}
    \mathcal{M}(\delta^\star):=\left\{(\mathbf{s}_1,\mathbf{s}_2,\mathbf{s}_3)\in\{-1,1\}^{3G}\;:\;\{d_{ij},d_{ik},d_{jk}\}=\delta^\star\right\},\label{paper3:eqn:solutions-set}
\end{equation}
with $i\neq{j}\neq{k}$, denote the set of microstates whose distances satisfy $\delta^\star$.
Since there is no external field, each spin marginal is uniform, hence $H(s_i)=G$ for all $i$.
The joint entropy over the ground-state ensemble $\mathcal{M}=\mathcal{M}(\delta^\star)$ is
\begin{equation*}
    H(\mathcal{M})=\log_2|\mathcal{M}|,
\end{equation*}
where $|\mathcal{M}|$ can be computed combinatorially (see Appendix A Sec.~\ref{paper3:sec:ground-state-degeneracy}).
To compute the entropy of a pair of spins, note that each pair $(\mathbf{s}_i, \mathbf{s}_j)$ can occur at any distance $d\in\delta^\star$ (i.e., there are $\binom{G}{d}$ choices).
Let $D$ be a random variable taking values in $\delta^\star$ with probabilities $p(D=d)=\mu_d/3$, where $\mu_d$ is the multiplicity of $d$ in $\delta^\star$.
Then
\begin{equation}
    H(\mathbf{s}_i,\mathbf{s}_j) = G + H(D) + \mathbb{E}_D\left[\log_2\binom{G}{D}\right],
    \label{paper3:eqn:pair_entropy}
\end{equation}
where $H(D)\in\left\{0,\,\log_2(3)-\tfrac23,\, \log_2(3)\right\}$ while 0 $\leq \log_2\binom{G}{D} \leq \log_2\binom{G}{\lfloor G/2\rfloor}$.
By combining these entropies into Eqs.~\eqref{paper3:eqn:u1-component}-\eqref{paper3:eqn:omega-decomposed}, we obtain
\begin{align}
    u_1(\mathbf{X}^3) &= 2G-H(\mathbf{s}_i,\mathbf{s}_j) \\
    u_2(\mathbf{X}^3) &= 2H(\mathbf{s}_i,\mathbf{s}_j) - G - \log_2|\mathcal{M}| \\
    \Omega(\mathbf{X}^3) & = u_1(\mathbf{X}^3) - u_2(\mathbf{X}^3).
\end{align}
These expressions make the mechanism clear.
Low-order dependencies ($u_1$) are controlled primarily by the pairs' combinatorial degeneracy due to typical pairwise distances, which is captured by $\mathbb{E}_D\left[\log_2\binom{G}{D}\right]$ (note, $H(D)$ is typically a comparatively small correction).
High-order dependencies ($u_2$) instead are large when the triplet is more constrained (small $|\mathcal{M}|$) than what pairwise statistics alone would predict, i.e., when $\log_2|\mathcal{M}|$ is small relative to the pairwise entropies ($H(\mathbf{s}_i,\mathbf{s}_j)$).
These measures neatly capture the key distinction between homophily and heterophily.

In the homophilous ground states, distances are pushed to the extremes ($d\in\{0, G\}$, see Table~I).
In this regime $\log_2\binom{G}{d}=0$, so $H(\mathbf{s}_i,\mathbf{s}_j)$ is reduced and $u_1$ is large.
This means that triadic constraints (small $\log_2|\mathcal{M}|$) coexist with strong pairwise constraints.
Informally, most of the information about the whole is disclosed at the pairwise level.
As a result, $u_1>u_2$ and $\Omega$ is strongly positive across $\alpha$.

In the heterophilous ground states, distances are pulled towards the threshold ($d\in\{ h, h+1\}$, see Table~I), where $\binom{G}{d}$ is largest.
Consequently, $H(\mathbf{s}_i,\mathbf{s}_j)\approx 2G$, indicating that pairs are almost independent.
However, the hypercube constraints still restrict the feasible triples of distances, so the ground-state manifold remains comparatively small (i.e., $\log_2|\mathcal{M}|$ does not simply scale as $3G$).
This is precisely the regime in which conditioning on a third variable reveals information that is invisible at the pairwise level: while $u_1$ is suppressed, $u_2$ remains relatively high, which is the hallmark of synergy-dominance.

\begin{figure}[H]
    \centering
    \includegraphics[width=1\linewidth]{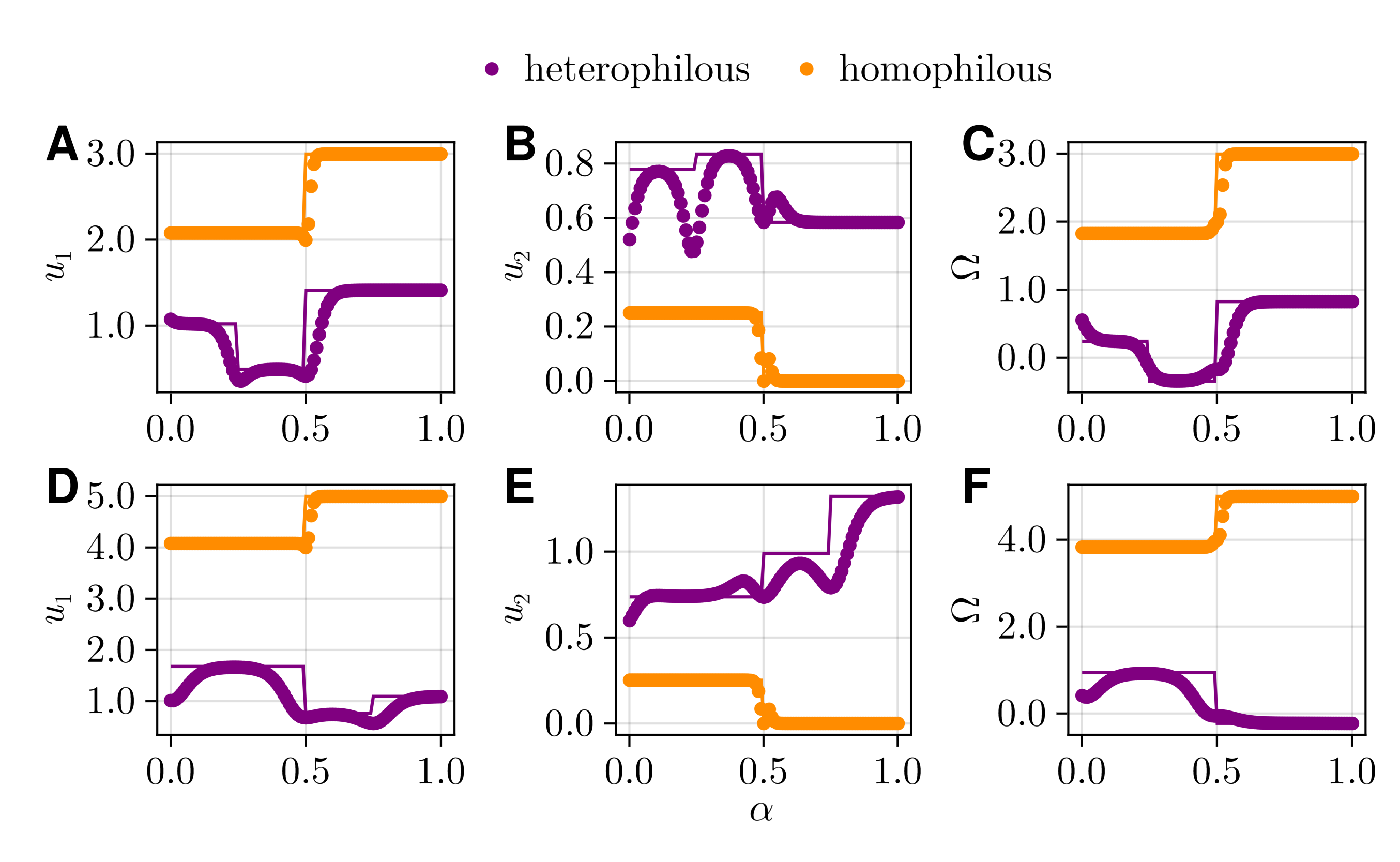}
    \caption[$u_1$, $u_2$ and $\Omega$ as a function of $\alpha$ in systems of size $N=3$.]{
    Homophily promotes strong pairwise dependencies ($u_1$) and redundancy-dominated organisation ($\Omega>0$), whereas heterophily suppresses $u_1$, enhances the relative contribution of $u_2$, and can induce synergy-dominance ($\Omega<0$).
    % $u_1$, $u_2$ and $\Omega$ as a function of $\alpha$ in systems of size $N=3$.
    Solid lines correspond to the ground-state predictions (zero-temperature), while markers correspond to the exact finite temperature values from the Boltzmann distribution ($\beta=20$).
    (\textbf{A}--\textbf{C}) show results for $G=3$.
    (\textbf{D}--\textbf{F}) show results for $G=5$.
    Finite temperature smooths the transitions across $\alpha$, but preserves the predicted informational regimes.
    At critical values of $\alpha$, two ground-state configurations coexist, hence the drops at finite temperatures.
    % For systems of size $N=3$, we show the low-order dependencies (\textbf{A}, \textbf{D}) and high-order dependencies (\textbf{B}, \textbf{E}) and the O-information (\textbf{C}, \textbf{F}) for varying $\alpha$.
    }
    \label{paper3:fig:static_O_info_exact_N3}
\end{figure}

In Fig.~\ref{paper3:fig:static_O_info_exact_N3} we compare the ground-state analytical predictions (continuous lines) with the exact computations from the Boltzmann distribution at finite $\beta$ (markers) for $G=3$ (such that $h$ is odd) and $G=5$ (such that $h$ is even).
At finite temperatures, the transitions as $\alpha$ varies are smoother than the analytical results, but preserve very well the overall regime (synergy- or redundancy-dominance) predicted by the ground-state analysis.
Homophilous systems display strong low-order dependencies ($u_1$) and consistent redundancy-dominated organisation ($\Omega>0$).
Heterophilous systems instead display consistently lower $u_1$ and higher $u_2$ than the homophily case.
When $G=3\;(\mathrm{mod}\;4)$, the system is synergy-dominated for intermediate values of $\alpha$.
When $G=1\;(\mathrm{mod}\;4)$, the system is synergy-dominated for $\alpha>0.5$.
In Appendix~D Sec.~\ref{paper3:sec:extended-IT-N3} we show the behaviour of small systems of size $N=3$ as $G$ increases.

\subsubsection{High-order interdependencies in large systems}\label{paper3:sec:IT-large-systems}
\begin{figure}
    \centering
    \includegraphics[width=1.0\linewidth]{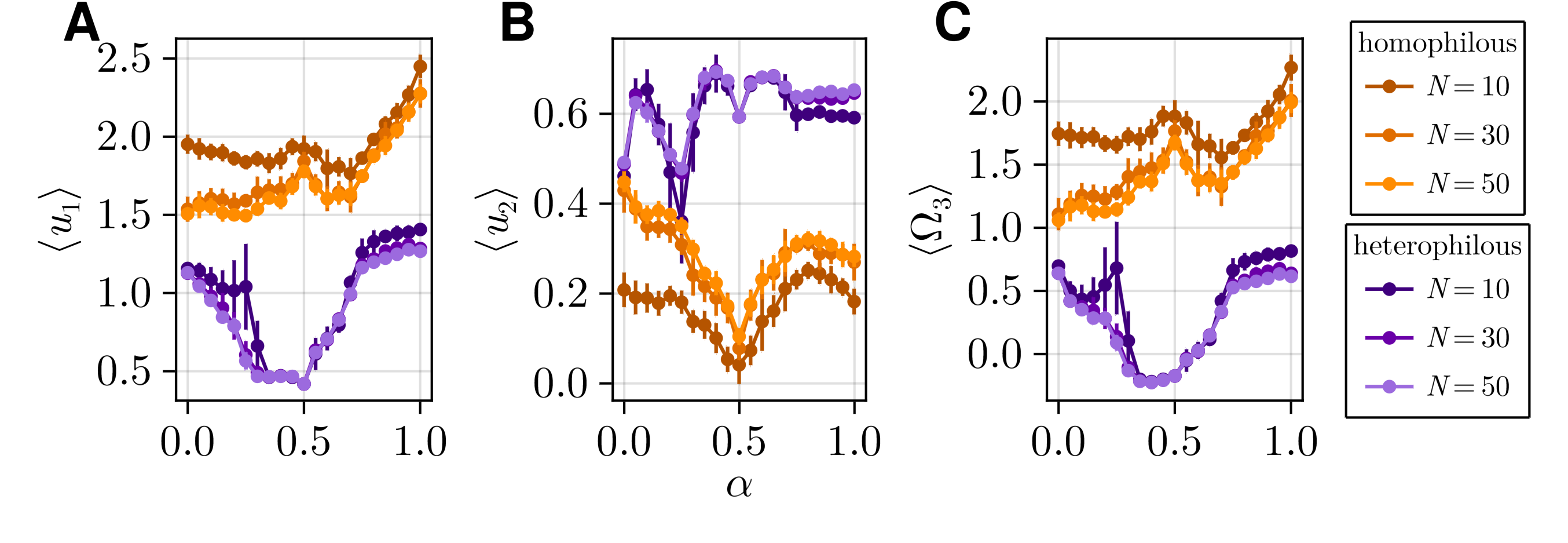}
    \caption[Triplet averaged $u_1$, $u_2$ and $\Omega_3$ across triplets as a function of $\alpha$ in large networks]{
        Large heterophilous systems display synergy-dominance for intermediate values of $\alpha$, consistent with $N=3$ analysis.
        Triplet averaged $u_1$ (\textbf{A}), $u_2$ (\textbf{B}) and $\Omega_3$ (\textbf{C}) across triplets as a function of $\alpha$ for systems with $G=3$, $\beta=20$, and using small-world graphs of size $N=10,\,30,\,50$ with average degree $k_{\mathrm{avg}}=5$ and probability of rewiring $\epsilon=0.2$.
        Error bars denote standard deviation of observables across $10$ independent replica ensembles.
        }
    \label{paper3:fig:O_info_large_N}
\end{figure}
The $N=3$ case isolates the mechanism in its minimal form.
We now ask whether the same informational architecture regimes persist when triangles are part of larger networks, where distinct triplets can share edges and constraints can thus propagate across the network.
Since the microstate space grows as $2^{NG}$, we limit our numerical analysis to high-order dependencies at the triplet level for $G=3$.

We quantify information about organisational structure by the triplet-averaged quantities
\begin{align}
\langle \Omega_3 \rangle \;&=\; \frac{1}{|\mathcal{T}|}\sum_{(i,j,k)\in\mathcal{T}} u_1(\mathbf{s}_i,\mathbf{s}_j,\mathbf{s}_k) - u_2(\mathbf{s}_i,\mathbf{s}_j,\mathbf{s}_k), \label{paper3:eqn:o-info-triplet-average}\\
\langle u_1 \rangle \;&=\; \frac{1}{|\mathcal{T}|}\sum_{(i,j,k)\in\mathcal{T}} u_1(\mathbf{s}_i,\mathbf{s}_j,\mathbf{s}_k) \label{paper3:eqn:u1-triplet-average} \\
\langle u_2 \rangle \;&=\; \frac{1}{|\mathcal{T}|}\sum_{(i,j,k)\in\mathcal{T}} u_2(\mathbf{s}_i,\mathbf{s}_j,\mathbf{s}_k) \label{paper3:eqn:u2-triplet-average}
\end{align}
where $\mathcal{T}$ is the set of connected triplets in the underlying graph $A$.
Note, all information-theoretic observables  are estimated from the replica ensemble.

In Fig.~\ref{paper3:fig:O_info_large_N}, we report estimates of $\langle\Omega_{3}\rangle, \langle u_1 \rangle$ and $\langle u_2 \rangle$ as a function of $\alpha$ on small-world graphs of sizes $N=10,\,30,\,50$ with fixed mean degree equal to $5$ and rewiring probability equal to $0.2$.
Despite the presence of many interacting triangles and low $G=3$, we recover the same qualitative regimes predicted by the $N=3$ analysis:
homophily yields consistently redundancy-dominated organisation ($\langle \Omega_3\rangle>0$),
whereas heterophily exhibits an intermediate range of $\alpha$ in which $\langle \Omega_3\rangle<0$, in great agreement with our analytical insights.
Thus, the synergistic behaviours observed under heterophily are not a peculiarity of the minimal system, but persist in larger, more structured networks where triadic constraints can overlap across connected triplets.
\subsection{Robustness analysis: local update scheme and parameter heterogeneities}\label{paper3:sec:robustness_analysis}

\begin{figure}[H]
    \centering
    \includegraphics[width=.95\linewidth]{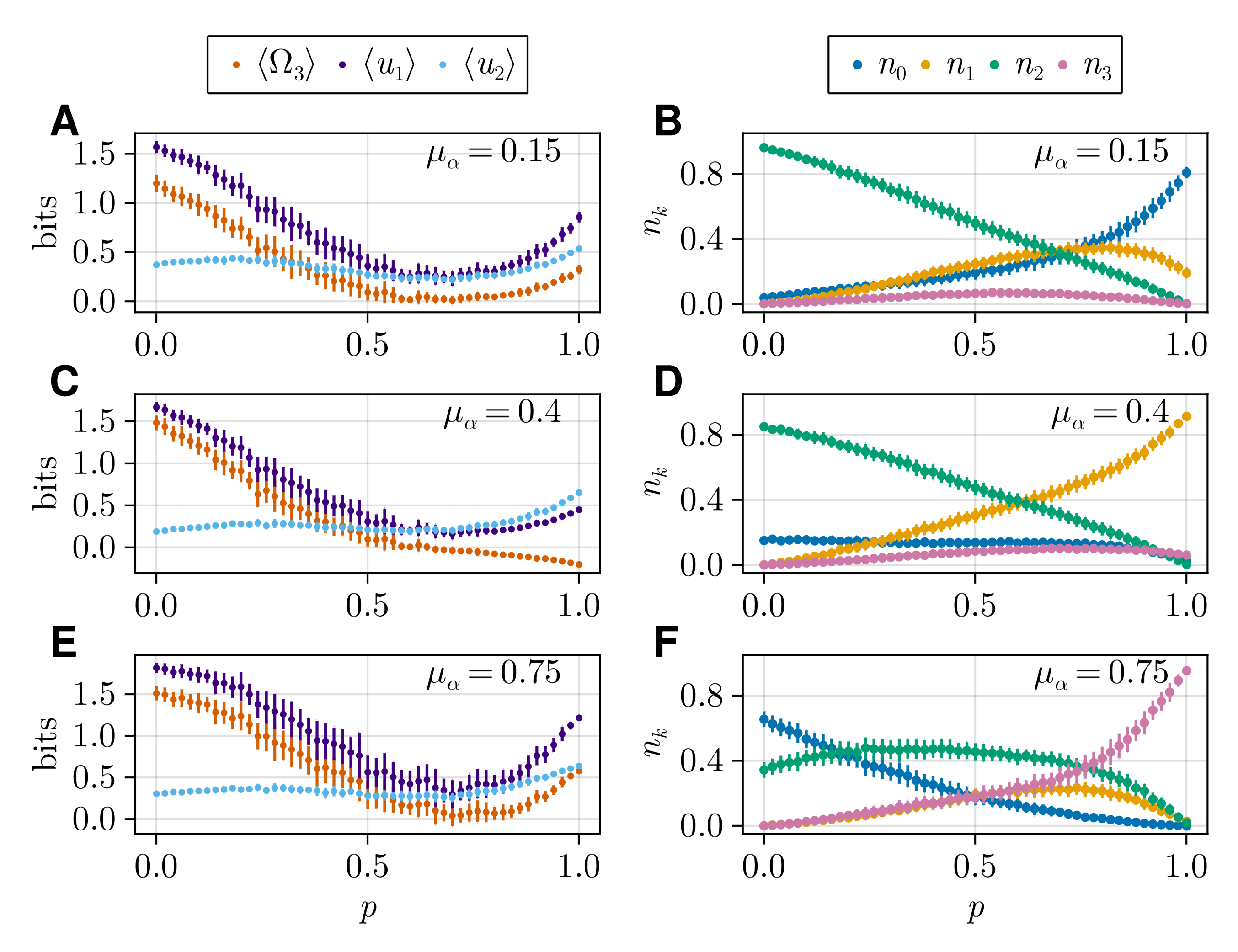}
    \caption[Robustness of the synergistic mechanism under mixed $\lambda_i$, heterogeneous $\alpha_i$, and local updates.]{Robustness of the synergistic mechanism under mixed $\lambda_i$, heterogeneous $\alpha_i$, and local updates.
    A fraction $p$ of nodes is heterophilous ($\lambda_i=1$), with the remainder homophilous ($\lambda_i=-1$).
    Rows correspond to $\mu_\alpha\in\{0.15,0.4,0.75\}$.
    (\textbf{A}, \textbf{C}, \textbf{E}) Show triplet-averaged $\langle{u_1}\rangle$, $\langle{u_2}\rangle$, and $\langle{\Omega_3}\rangle$ as a function of $p$ for different $\mu_\alpha$.
    (\textbf{B}, \textbf{D}, \textbf{F}) Show motif densities $n_k$ as a function of $p$ for different $\mu_\alpha$.
    % Across all regimes, increasing $p$ suppresses $\langle{u_1}\rangle$ while maintaining $\langle{u_2}\rangle$, and reorganizes triangle patterns of interaction.
    Systems of size $N=50$, with $G=3$, on small-world graphs mean degree $k_\mathrm{avg}=5$, rewiring probability $\epsilon=0.2$.
    Error bars denote standard deviation of observables across $10$ independent replica ensembles.}
    \label{paper3:fig:robustness}
\end{figure}

In the previous sections, we have identified a precise mechanism for the emergence of synergies under the heterophilous adaptive rule.
Heterophily suppresses low-order dependencies (small $u_1$) while maintaining high-order dependencies by favouring particular triadic configurations that can't be predicted by local (pairwise) information only (high $u_2$).
We now test whether this twofold mechanism persists under three departures from the idealised symmetric case:
($i$) using mixed populations, where both heterophilous ($\lambda_i=1$) and homophilous ($\lambda_i=-1$) elements are present,
($ii$) using heterogeneous pressures $\alpha_i$,
($iii$) and adopting a local update rule, whose transition probabilities $p(S\to S')$, see Eq.~\eqref{paper3:eqn:transition-probability}, depend on local energy changes $\Delta E^{(i)}= E^{(i)}(S') - E^{(i)}(S)$, which does not guarantee detailed balance.
Specifically, we analyse systems of size $N=50$ on small world graphs (average degree $5$ and probability of rewiring $0.2$) and vary the proportion $p\in[0,1]$ of heterophilous elements in the system, with the remaining fraction of nodes being homophilous.
To model heterogeneous pressures, we draw $\alpha_i\sim\mathcal{N}(\mu_\alpha, \sigma_\alpha)\in[0,1]$.

With these settings, the homophilous and heterophilous symmetric systems correspond to systems with $p=0$ and $p=1$, respectively.
Since both of these symmetric systems undergo phase transitions as $\alpha$ varies (see Fig.~\ref{paper3:fig:triangles_large_N_example} and Table~$1$),
we study all possible cases for $G=3$ by analysing systems with $\mu_\alpha\in\{0.15, \,0.4,\,0.75\}$ and fixed $\sigma_\alpha=0.05$.
For each $\mu_\alpha$, we report in Fig.~\ref{paper3:fig:robustness}~(\textbf{A}, \textbf{C}, \textbf{E}) the O-information, $u_1$, and $u_2$, averaged across triplets (Eqs.~\eqref{paper3:eqn:o-info-triplet-average}-\eqref{paper3:eqn:u2-triplet-average}).

Fig.~\ref{paper3:fig:robustness}~(\textbf{A}, \textbf{C}, \textbf{E}) shows that introducing heterophilous elements leads to a rapid decrease of low-order dependencies: $\langle{u_1}\rangle$ decreases sharply as $p$ starts to increase.
In contrast, $\langle{u_2}\rangle$ remains consistently non-zero, and displays a non-monotonic behaviour.
Crucially, for $\mu_\alpha=0.4$ and $p>0.5$, $\langle{\Omega_3}\rangle$ becomes negative (synergy-dominance regime), in line with our theoretical predictions.
For $\mu_\alpha=0.15$ and $\mu_\alpha=0.75$ instead, $\langle{\Omega_3}\rangle$ decreases as $p$ increases, but remains positive.
Notably, $\langle{u_2}\rangle$ is non-zero even for $p=0$ (symmetric homophilous systems).
This does not imply synergy-dominance: $\langle{\Omega_3}\rangle$ remains positive because $\langle{u_1}\rangle$ is much larger than $\langle{u_2}\rangle$.
However, it suggests that parameter heterogeneity and local updates induce triadic dependencies beyond what is captured by pairwise marginals, while homophily continues to enforce strong pairwise alignment.

Results shown in Fig.~\ref{paper3:fig:robustness}~(\textbf{B}, \textbf{D}, \textbf{F}) demonstrate that patterns of triadic motifs are reorganized in a controlled way as $p$ increases.
At the extremes $p=0$ and $p=1$, the dominant triadic pattern is consistent with our analytical predictions from Sec.~\ref{paper3:sec:triadic_constraints}.
For instance, consider the case for $\mu_\alpha = 0.4$.
When $p\ll 1$ the system mostly contains homophilous elements and the dominant interaction pattern is $\Delta_2$.
As the number of heterophilous elements increases, the system approaches the $\Delta_1$-dominated regime, consistent with the regimes identified in Table~1 and Fig.~\ref{paper3:fig:triangles_large_N_example}.
Thus, increasing $p$ does not merely add noise in the organisation: it produces a systematic reconfiguration of triadic motifs and preserves non-trivial high-order dependencies ($u_2>0$).

Overall, these analyses demonstrate that the core mechanism we studied in the previous section for the idealized case strongly persists when parameter heterogeneities and local update schemes are introduced.
Next, we show how our analytical insights are directly applicable to study dynamical opinion formation.
\subsection{Case Study: Disrupting Polarization}\label{paper3:sec:disrupting_polarization}
The homophilous symmetric model has recently been used to study a wide range of phenomena in computational social science: the emergence of balance~\cite{Pham2022homophily,Galesic2025experimentHomophily}, self-assemblies~\cite{Korbel2023selfAssembly}, and polarization~\cite{Thurner2025polarization}.
In this framework, nodes in the network represent individuals in a social system, each having $G$ binary opinions $\mathbf{s}_i$.
Positive couplings $J_{ij}$ represent friendly relationships, while negative ones represent animosity.
Individuals interact with each other according to Eq.~\eqref{paper3:eqn:local_hamiltonian}, which can be interpreted as a measure of local cognitive dissonance~\cite{Heider1946SocialBalance}, that individuals try to minimize by changing their opinion.
Homophilous individuals change their opinions to align more closely to their friends, while moving away from the opinion of their enemies.
Here, we interpret heterophily as the opposite tendency: individuals update their opinions by moving toward their enemies (a conciliatory tendency) and away from their friends (a contrarian tendency~\cite{Galam2004contrarian}).
In the homophilous case, when $\alpha>0.5$, the system reaches a state in which all individuals are friends with each other ($\Delta_0$-dominated regime or consensus regime).
When $\alpha<0.5$ instead, the system is in a $\Delta_2$-dominated regime (polarized regime), characterized by the emergence of $2$ or more antagonistic subgroups~\cite{Thurner2025polarization,davis1967clusteringSBT}.
Here we focus on the polarized scenario and, through the lens of information-theory, ask how introducing heterophilous individuals alters the dynamics of opinions in a polarized society.

We conduct the following experiment.
Using $G=3$, we start by simulating a homophilous system with heterogeneous $\alpha_i\sim\mathcal{N}(0.4,\,0.05)$ for $t_0=10^4$ Metropolis sweeps (each sweep being $N$ updates).
This ensures that polarization is reached (see case for $\mu_\alpha=0.4$ and $p=0$ in Fig.~\ref{paper3:fig:robustness}).
As in~\cite{Thurner2025polarization}, we measure polarization as the variance in the overlaps between individuals,
\begin{equation}
    \psi = \mathrm{Var}(\{O_{ij}\}_{i<j})
\end{equation}
across all pairs $i,j$ of individuals.
After this initial phase, we perturb the system by turning a fraction $p$ of elements from homophilous to heterophilous, i.e., $\lambda_i=-1\to\lambda_i=1$.
For each value of $p$, we construct $50$ replica ensembles of size $R=10^5$ and record snapshots of the systems state $S(t_b),\,S(t_b+1)$ at times $t_b=0,5,10,\,\dots,\,50$ after the perturbation.
Each replica within an ensemble is initialized from the same polarized configuration at $t_b=0$.
Thus, replicas diverge only due to stochastic updates after the perturbation.
For this experiment, as in~\cite{Thurner2025polarization}, we use $\beta=3.4$, and small-world networks with average degree $4$ and probability of rewiring equal to $0.175$.

Additionally, to facilitate direct comparison with Refs.~\cite{Pham2022homophily,Thurner2025polarization}, in this section we adopt the same local Metropolis update scheme used there.
Specifically, after proposing a single-site flip at node $i$, the move is accepted with probability
\begin{equation*}
    p(S\to S') = \mathrm{min}\left\{1,\exp(-\beta\Delta E^{i})\right\}
\end{equation*}
where $\Delta E^{i}$ is the change in local energy.
Compared to the update rule defined in Sec.~\ref{paper3:sec:methods} this acceptance rule uses local, rather than global, energy changes, and, unlike the local Glauber updates used in Sec~\ref{paper3:sec:robustness_analysis}, here energy-lowering moves are always accepted.

\begin{figure}
    \centering
    \includegraphics[width=.99\linewidth]{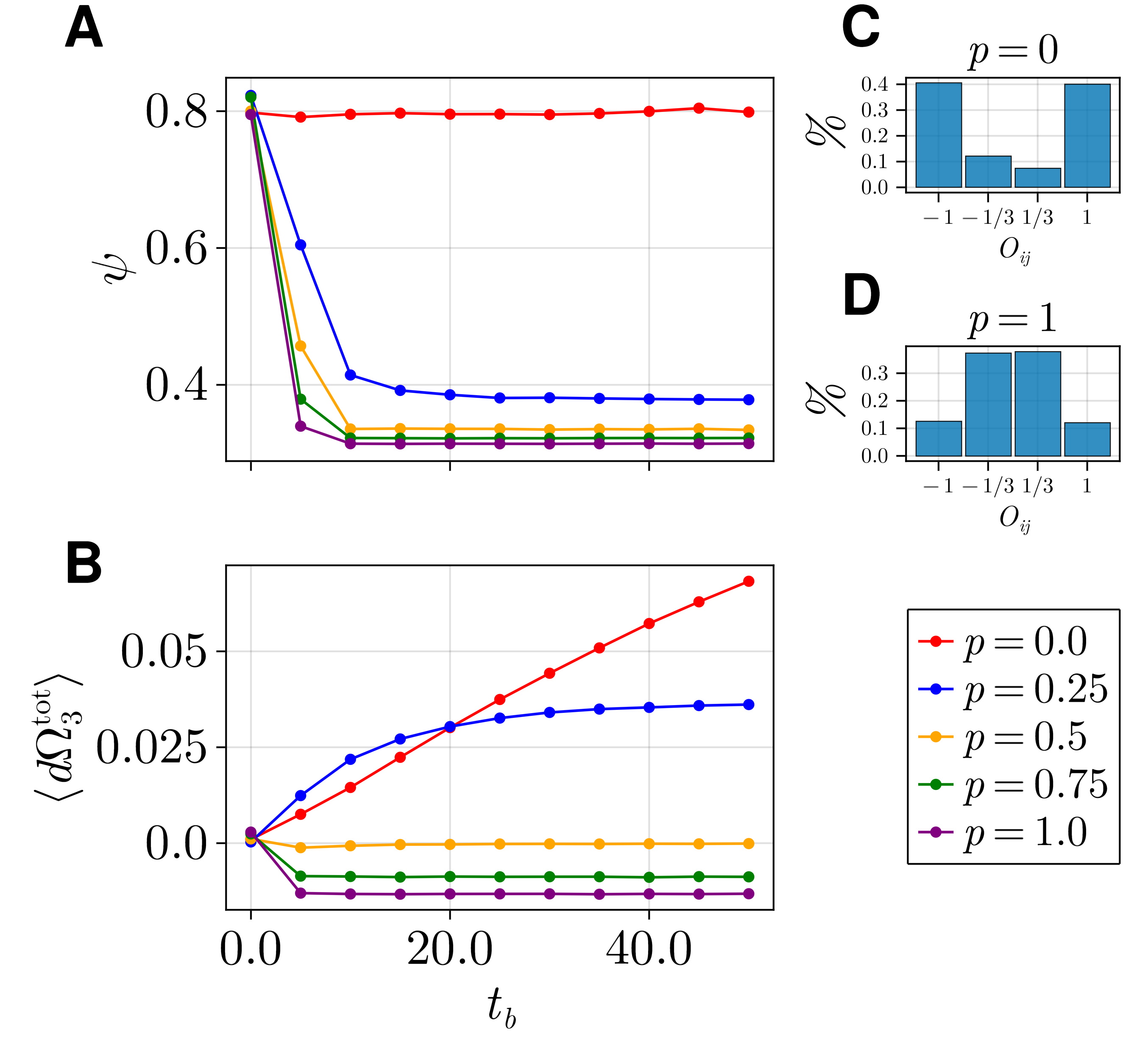}
    \caption[Heterophily disrupts polarization and induces synergy-dominated dynamical information sharing]{Heterophily disrupts polarization and induces synergy-dominated dynamical information sharing.
    (\textbf{A}) Polarization $\psi$ as a function of $t_b$ for different values of $p$.
    (\textbf{B}) Total dynamical O-information $\langle d\Omega^{\mathrm{tot}}_3\rangle$ as a function of $t_b$ for different values of $p$.
    (\textbf{C}) Distribution of pairwise overlaps $O_{ij}$ in the homophilous case with $p=0$.
    (\textbf{D}) Distribution of pairwise overlaps $O_{ij}$ in the heterophilous case $p=1$ (after perturbation).
    Systems of size $N=20$ with $G=3$ on small-world graphs with mean degree $k_\mathrm{avg}=4$, rewiring probability $\epsilon=0.175$.
    As in~\cite{Thurner2025polarization} we use $\beta=3.4$.
    Each value is the mean across 50 runs.
    The corresponding individual trajectories are reported in Fig.~\ref{paper3:fig:polarization-extra}.
    }
    \label{paper3:fig:perturb-polarization}
\end{figure}

Fig.~\ref{paper3:fig:perturb-polarization} shows our results for systems of size $N=20$.
Polarization in the system is reduced for any fraction $p$ of individuals that turn heterophilous (Fig.~\ref{paper3:fig:perturb-polarization} A).
For $p\geq0.5$, $\psi$ reaches values close to $\approx1/3$, indicating that opinions (binary vector of size $3$) are well-mixed.
For $p=0$ instead, the system remains highly polarized with $\psi\approx0.8$, indicating that there are two factions of individuals with strongly opposed opinions.
This is consistent with heterophily promoting the minimization of the overlaps (equivalently distances $d\in\{h,\,h+1\}$), and homophily promoting the maximisation of overlaps ($d\in\{0,\,G\}$).
In Fig.~\ref{paper3:fig:perturb-polarization}~\textbf{C} and~\textbf{D} we confirm these observations by showing the histogram of the overlaps $O_{ij}$ for the $p=0$ and $p=1$ cases.

We then use the total dynamical O-information to study how the perturbation alters dynamical high-order dependencies over time.
Here, $d\Omega^{\mathrm{tot}}_3>0$ indicates that, after the perturbation, an individual’s next opinion is dominated by redundant information.
Conversely, $d\Omega^{\mathrm{tot}}_3<0$ indicates that an individual’s next opinion is better predicted by synergistic information, which only becomes available when the two neighbours are considered jointly, while conditioning on the individual’s own past state ($\tau=1$).
By evaluating $d\Omega^{\mathrm{tot}}_3(t_b)$ at successive times $t_b$ after the perturbation, we obtain an explicitly time-dependent measure of the system's evolving dynamical high-order dependencies.
In contrast with our previous results using the O-information, here we can't directly compare numerical result to the corresponding analytical calculations.
Thus, we report control-corrected estimates $d\Omega^{\mathrm{tot}}_3 - d\Omega^{\mathrm{tot}}_{3,\mathrm{shuf}}$ where $d\Omega^{\mathrm{tot}}_{3,\mathrm{shuf}}$ is computed from time-shuffled data.
Fig.~\ref{paper3:fig:perturb-polarization} (\textbf{B}) shows that introducing heterophilous individuals drives a shift in the dynamical informational architecture.
When the majority of individuals are homophilous (i.e., $p<0.5$), the $\langle{d\Omega_3^\mathrm{tot}}\rangle$ remains positive.
When heterophilous individuals are the majority, $\langle{d\Omega_3^\mathrm{tot}}\rangle$ quickly becomes negative.
In Appendix~C Fig.~\ref{paper3:fig:polarization-extra} we report the individual trajectories underlying the averages shown in Fig.~\ref{paper3:fig:perturb-polarization}~(\textbf{B}).

Taken together, Fig.~\ref{paper3:fig:perturb-polarization}~\textbf{A}–\textbf{B} show that heterophily both reduces polarization and induces synergy-dominated information dynamics, providing a concrete example of the general mechanism identified in Sec.~\ref{paper3:sec:IT_analysis}.
Importantly, this is not a trivial extension of the equal-time results.
A negative (equal-time) O-information does not, in general, imply negative total dynamical O-information.
The observed negative values of $d\Omega_3^\mathrm{tot}$ thus indicate that heterophily reshapes not only the instantaneous informational architecture of the system, but also the collective mechanisms by which opinions evolve over time.
% discussion
\section{Discussion}
In this work, we have identified a parsimonious route by which synergistic interdependencies can emerge in a self-organized manner in systems governed by pairwise adaptive rules. 
Heterophily can lead to regimes in which the informational architecture is irreducible to pairwise statistics, as quantified by the equal-time O-information and the total dynamical O-information.
Conversely, we have shown that homophily yields redundancy-dominated organisations.

In analytically tractable systems of size $N=3$, we uncover a precise twofold mechanism:
($i$) At the pairwise level, heterophily pulls distances (which quantify alignments) towards the similarity threshold $d\in\{h,h+1\}$.
This maximises pairwise degeneracy (maximises $\log_2\binom{G}{d}$), thus favouring near pairwise independence within each pair (suppressing $I(X_i;X_j)$) and reducing low-order dependencies ($u_1$).
($ii$) Despite weak pairwise dependencies, heterophily does not lead to random configurations without any structure;
triadic dependencies persist because not every triple $\delta$ can be realized. 
Heterophily therefore selects a particular subset of microstate configurations.
Crucially, this subset cannot be reconstructed from pairwise mutual informations alone, and the resulting organisation maintains comparatively large high-order dependencies ($u_2$).

Homophily exploits the same geometric mechanism, but promotes redundant dependencies instead.
This is particularly clear for the case of $\alpha<0.5$ where pairs prefer maximal dissimilarity.
The ``ideal'' configuration in which all pairs are maximally dissimilar is geometrically impossible.
Therefore, homophily selects microstate configurations $\delta=\{0,G,G\}$ (corresponding to $\Delta_2$) that provide the smallest deviation from this ideal configuration.
In these states, pairwise mutual information is maximised, yielding a redundancy-dominated regime: locally available information suffices to predict the state of each element.

A simple intuition is the following.
Fix a random binary string of length $G$ as the state of node $i$.
In the heterophilous regime, neighbours $j$ are obtained by flipping approximately half the spins.
This generates many combinatorially distinct microstates with similar pairwise distances ($d\in\{h,h+1\}$).
In this case, knowledge of the $(i,j)$ pair relationship discloses limited information about a third element $k$ connected to both $i$ and $j$.
In contrast, in the homophilous case, $j$ is either approximately copying $i$ or flipping nearly all spins.
In this case, pairwise degeneracy is significantly reduced, and knowledge of the $(i,j)$ pair relationship discloses all (or almost all) the information needed to infer the state of a third element $k$.

In Sections~\ref{paper3:sec:triangle-analysis-large}, \ref{paper3:sec:IT-large-systems} and \ref{paper3:sec:robustness_analysis}, we showed 
that these geometric constraints manifest structurally as over-represented triangle motifs, and informationally as a shift in the balance between $u_1$ and $u_2$ (averaged across triplets).
First, large systems self-organize into particular configurations in which the same triangles identified for $N=3$ are over-represented (Sec.~\ref{paper3:sec:triangle-analysis-large}, Fig.~\ref{paper3:fig:triangles_large_N_example}).
This provides a precise, geometric explanation for the emergence of balance in opinion-dynamics models with state-dependent coupling signs ($J_{ij}$)~\cite{Pham2022homophily,Thurner2025polarization}.
The $N=3$ ground-state solutions predict the transition from $\Delta_2$-dominance (polarized) to $\Delta_0$-dominance (consensus) as $\alpha$ increases, in agreement with the motif transition reported in~\cite{Pham2022homophily}.
Second, in Sec.~\ref{paper3:sec:IT-large-systems} and~\ref{paper3:sec:robustness_analysis} the same organisation of information studied for small systems persists qualitatively in large systems under parameter heterogeneities and local update schemes.
Homophilous systems lead to balanced structures in which the organisation of information (at the triplet level) is highly redundant.
In contrast, heterophily can lead to both balanced and antibalanced structures in which pairwise dependencies are suppressed while high-order dependencies persist (Fig.~\ref{paper3:fig:O_info_large_N}).
For example, for $G=3$ and $0.25<\alpha<0.5$, the antibalanced regime we observe is synergy-dominated, in great agreement with our theory for $N=3$.

The emergence of balanced and antibalanced patterns of interactions has important consequences for the macroscopic organisation of social systems~\cite{Heider1946SocialBalance,Cartwright1956,Estrada2019rethinkSBT,Marvel2009energySBT,Antal2005SBTdynamics,Facchetti2011SBTlargeSocial, Marvel2011SBT,Kirkley2019SBT,Healy1973balanceInternational,Altafini2012SBT,Gallo2024SBTheterogeneous}.
For example, a core result from structural (or social) balance theory is that networks with predominantly balanced triangles naturally form two polarized modules: individuals within the same module are positively connected, while individuals between modules are negatively connected~\cite{Cartwright1956}.
In this framework, a recent line of work has shown that many collective properties of social systems can emerge from pairwise interactions, using the same homophilous model we studied here~\cite{Pham2022homophily,Thurner2025polarization,Korbel2023selfAssembly} and supported by experimental evidence~\cite{Galesic2025experimentHomophily}.

To show a concrete application of the mechanisms we have uncovered here, we considered the effect of introducing heterophilous elements to a polarized society and studied, through the lens of information theory, how opinions form after the system is perturbed.
Heterophily has been implemented in multiple ways in the opinion-dynamics literature, and in several models it has been shown to promote consensus~\cite{Motsch2014heterophily,Wu2024heterophily}.
Here we study a different setting: a pairwise, spin-glass–like model with adaptive signed couplings, which, to the best of our knowledge, has not been examined in this form and, importantly, not through the lens of information-theory.
More specifically, we measured polarization after introducing heterophilous elements, together with the total dynamical O-information.
This dynamical information-theoretic measure provides important insights into how opinions are formed.
When negative, $d\Omega_3^\mathrm{tot}$ indicates that the whole (groups of three individuals) predicts the state of an individual better than any single neighbour in isolation.
When positive instead, $d\Omega_3^\mathrm{tot}$ indicates that an individual's opinion mostly comes from redundant information shared by the individuals in the group.

Our results show that heterophilous individuals not only reduce polarization, but also induce a dynamical regime in which group-level mechanisms play a stronger role than pairwise alignment in shaping opinions.
These results provide a concrete, dynamical example of the general mechanism identified in Sec.~\ref{paper3:sec:IT_analysis}:
heterophily weakens pairwise dependencies while preserving high-order ones, and this combination can both disrupt polarized configurations and can lead to the emergence of synergy-dominated information dynamics.
This initial analysis naturally points to numerous directions for future work, to analyse larger systems, more fine grained information theoretic measures, and conduct more experiments and comparisons with empirical data.

Overall, together with our work on antibalance in Gaussian systems~\cite{Caprioglio2026synergisticMotifs}, the results we presented here point to a more general mechanistic principle for the emergence of synergy in dyadic systems:
when pairwise alignment is suppressed, but the system remains subject to non-trivial global constraints, high-order dependencies can emerge and dominate over low-order ones.
In~\cite{Caprioglio2026synergisticMotifs}, these constraints are imposed by the signed interaction topology.
Here, they arise from the interplay between geometric realizability constraints of the state space and adaptive dynamics.
In natural and artificial adaptive systems this mechanism may be realized whenever units are driven to differentiate under shared constraints, 
for example through niche differentiation, or disassortative mate choice in ecology or through multimodal learning in recurrent networks, where the objective function and the statistical structure of the training data may act as shared global constraints while different units are driven to specialize.

\subsection*{Limitations and future research directions}\label{paper3:sec:limitations}
The mechanisms and behaviours we studied depend on the discrete geometry of the attraction/repulsion model and on the fact that realizability constraints on the hypercube of dimension $G$ are non-trivial.
Whether similar mechanism and behaviour could persist in continuous models, such as oscillatory networks, is an interesting open question.

While the $N=3$ model, at zero-temperature, is solved exactly for arbitrary $G$, large-$N$ simulations are numerical and necessarily focus on small values of $G$ due to the rapidly expanding state space.
An interesting direction for future studies is to fully characterize how the synergy window in $\alpha$ scales with the system size $N$, how it depends on the average degree of the network, clustering coefficient, modularity, the presence of hubs etc. and how quickly the large-$N$ behaviour approaches asymptotic regimes.

Additionally, we quantified multivariate organisation through the average triplet O-information.
This allows us to specifically capture the distinction between pairwise and triadic interdependencies.
However, using only the triplet O-information we can't ultimately determine how interdependencies are organized across larger groups (i.e., quadruplets and beyond).

Finally, our model uses binary spins, a fixed interaction substrate $A$, and a simple distance-based, state-dependent coupling sign.
Real systems may involve weighted and directed interactions, external fields, and adaptive rewiring of the interactions.

\bibliography{references}

\clearpage

\appendix

\onecolumngrid

\section*{Supplementary Information}
\renewcommand{\thefigure}{S\arabic{figure}}
\setcounter{figure}{0}
\renewcommand{\theequation}{S\arabic{equation}}
\setcounter{equation}{0}

\section{Additional Technical Details}
\subsection{Ground States Degeneracy}\label{paper3:sec:ground-state-degeneracy}
To start, consider the fixed tuple $\delta'=(d_{ij},d_{ik},d_{jk})$ (i.e., without permutation of the indeces $i,j,k$).
We wish to find the cardinality of $\mathcal{M}(\delta')$.
Without loss of generality, fix $s_1=x=(x^1, x^2, x^3, \dots)$ for any odd $G$.
At each site $g\in[1,2,3,\dots,G]$ we have the following $4$ cases:
\begin{itemize}
    \item type $0$: where $s_2^g = x^g$ and $s_3^g = x^g$,
    \item type $2$: where $s_2^g \neq x^g$ and $s_3^g = x^g$,
    \item type $3$: where $s_2^g = x^g$ and $s_3^g \neq x^g$,
    \item type $23$: where $s_2^g \neq x^g$ and $s_3^g \neq x^g$.
\end{itemize}
Let $t_p$ the number of sites of type $p$.
Since for each site $g$ we must have exactly one of these cases, we have $t_0+t_2+t_3+t_{23}=G$.
Note that $d_{12}$ must be equal to the number of times a site $g$ is of type $2$ plus the number of times a site $g$ is of type $23$.
In general, we can write $d_{12}=t_2+t_{23}$,\;$d_{13}=t_3+t_{23}$ and $d_{23}=t_2+t_3$, allowing us to obtain a closed form for each $t_p$:
\begin{align*}
    t_2 &= \frac{d_{12}+d_{23}-d_{13}}{2} \\
    t_3 &= \frac{d_{13}+d_{23}-d_{12}}{2} \\
    t_{23} &= \frac{d_{12}+d_{13}-d_{23}}{2} \\
    t_0 &= G - \frac{d_{12}+d_{23}+d_{13}}{2}.
\end{align*}
To obtain $|\mathcal{M}(\delta')|$, we proceed as follows:
($i$) choose the $t_{23}$ sites in which both $s_2$ and $s_3$ differ from $s_1$,
($ii$) from the remaining sites, choose the $t_2$ sites in which only $s_2$ differs from $s_1$,
($iii$) from the remaining sites, choose the $t_3$ sites in which only $s_3$ differs from $s_1$.
The remaining $t_0$ sites are then automatically assigned.
More concisely, we can then write
\begin{equation}
    |\mathcal{M}(\delta')| = 2^G \binom{G}{t_{23}} \binom{G-t_{23}}{t_2} \binom{G-t_{23}-t_2}{t_3} = 2^G\frac{G!}{t_0!t_2!t_3!t_{23}!}.\label{paper3:eqn:degeneracy}
\end{equation}
Now, consider $\delta=\{d_{ij},d_{ik},d_{jk}\}$, which allows the permutation of the indeces $i,j,k$.
We obtain $|\mathcal{M}(\delta)|$ by multiplying $|\mathcal{M}(\delta')|$ by the number of permutations of the distances in $\delta$.
We only have three cases:
($i$) if all distances are equal, the number of permutations is just $1$,
($ii$) if only two distances are equal, the number of permutations is $3$,
($iii$) if all distances are different, the number of permutations is $6$.

\textbf{Example.} Consider the case for $\{0,G,G\}$ and $G=3$.
In this case, by fixing the distances, we have $t_2=t_3=0$, $t_{23}=3$, and $t_0=0$, such that $|\mathcal{M}((0,3,3))|=8$.
Then, since there are $3$ permutations of $\{0,3,3\}$, the number of degenerate ground states is $|\mathcal{M}(\{0,3,3\})|=24$ and $H(|\mathcal{M}|)\approx4.6$ bits.

\subsection{Mixed homophily and heterophily for \texorpdfstring{$N=3$}{N=3}}
\label{paper3:sec:heterogeneous-lambda}
Let us consider the case for mixed values of $\lambda_i$ for systems of size $N=3$.
We have two options for $(\lambda_1, \lambda_2, \lambda_3)$:
either choose two negative $\lambda_i$ and one positive or two positive $\lambda_i$ and one negative.
Since Eq.~\eqref{paper3:eqn:global_hamiltonian} assumes $\lambda_i=\lambda$ for all $i$, the global energy must instead be written as
\begin{equation}
    E(\delta) = \frac1G\sum_{i<j}(\lambda_i+\lambda_j)f_{\alpha,G}(d_{ij}).
   \label{paper3:eqn:global_hamiltonian_heterogeneous} 
\end{equation}

Let us consider first the case for two negative $\lambda_i$ (i.e., two homophilous elements and only one heterophilous).
If $\alpha_i=\alpha$ for all $i$, any permutation of the elements $(\lambda_1, \lambda_2, \lambda_3)$ does not change the global energy.
Without loss of generality, assume that $\lambda_1=1$ while $\lambda_2=\lambda_3=-1$.
Then, from Eq.~\eqref{paper3:eqn:global_hamiltonian_heterogeneous} two terms cancel, and we are left with $E=-\tfrac2Gf_{\alpha,G}(d_{23})$.
Since the global energy is negative, we maximise $f_{\alpha,G}(d_{23})$ by choosing $d_{23}\in\{0,G\}$, depending on the value of $\alpha$.
Using the conditions Eqs.~\eqref{paper3:eqn:parity}--\eqref{paper3:eqn:triangle} we can restrict the choice of the two other distances to $d_{12}=x$ and $d_{13}=G-x$ with $x\in[0,G]$.
More compactly, we can write the solutions as:
\[
    E=
        \begin{cases}
            E_{\Delta_2}=E_{(x,G-x,G)}=-2(1-\alpha) & \text{if }\alpha\leq1/2, \\
            E_{\Delta_0} = E_{\Delta_2} = E_{(x,x,0)} = -2\alpha & \text{if }\alpha\geq1/2.
        \end{cases}
\]
Thus, in this case only balanced solutions are allowed: if $\alpha<1/2$ only balanced ground states with two negative edges are allowed ($\Delta_2$), while for $\alpha>1/2$ we can have either $\Delta_0$ or $\Delta_2$ depending on whether $x\leq{h}$. 

For the case with two heterophilous elements and only one homophilous we quickly find that (assuming $\lambda_1=\lambda_2=1$ and $\lambda_3=-1$) $E=\tfrac2Gf_{\alpha,G}(d_{12})$ such that $d_{12}\in\{h,h+1\}$.
If $\alpha<1/2$ then $d_{12}=h$ minimizes the energy, obtaining $E=2\alpha$, while for $\alpha>1/2$ we have that $d_{12}=h+1$ minimizes the energy, and therefore $E=2(1-\alpha)$.
In both cases, depending on the parity of $h$, all possible triangles $\Delta_k$ are possible except for $k=3$.

Finally, note that in these cases for heterogeneous $\lambda_i$, the attraction-repulsion tendencies select only one pairwise distance while leaving the other two symmetrically unconstrained.
Because of this symmetry, there are no synergistic nor redundant interdependencies and thus $\Omega=0$.
\newpage
\section{Sensitivity Analyses}\label{paper3:sec:sensitivity-analyses}
\subsection[Sensitivity against burn-in time]{Sensitivity against burn-in time $t_b$}
\begin{figure}[H]
    \centering
    \includegraphics[width=.9\linewidth]{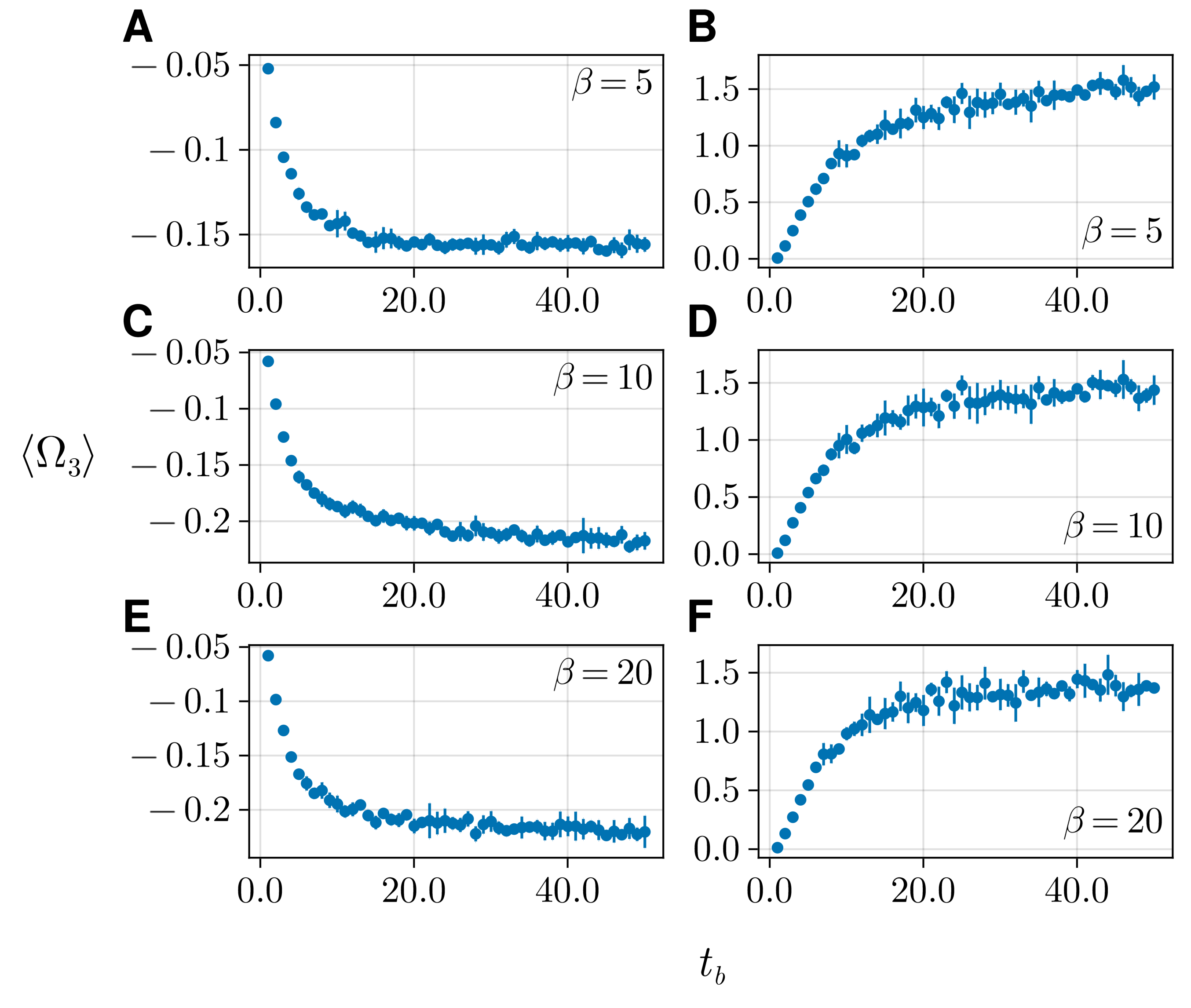}
    \caption[Sensitivity analysis with respect to burn-in time $t_b$]{
        Sensitivity analysis with respect to burn-in time $t_b$ for heterophilous systems of size $N=30$ and fixed $\alpha=0.4$.
        Left column (\textbf{A}, \textbf{C}, \textbf{E}) heterophilous systems.
        Right column (\textbf{B}, \textbf{D}, \textbf{F}) homophilous systems.
        Rows correspond to $\beta=5.0$ (\textbf{A}, \textbf{B}).
        Middle row correspond to $\beta=10.0$ (\textbf{C}, \textbf{D}).
        Bottom row correspond to $\beta=20.0$ (\textbf{E}, \textbf{F}).
        The results shown for $\alpha=0.4$ exemplify the behaviour observed for other values of $\alpha$.
        Each value is the mean across 5 replica ensembles each of size $R=10^5$, and error bars denote standard deviation.
    }
    \label{paper3:fig:sensitivity-t-record}
\end{figure}

\clearpage
\subsection[Sensitivity against inverse temperature]{Sensitivity against inverse temperature}
\begin{figure}[H]
    \centering
    \includegraphics[width=.9\linewidth]{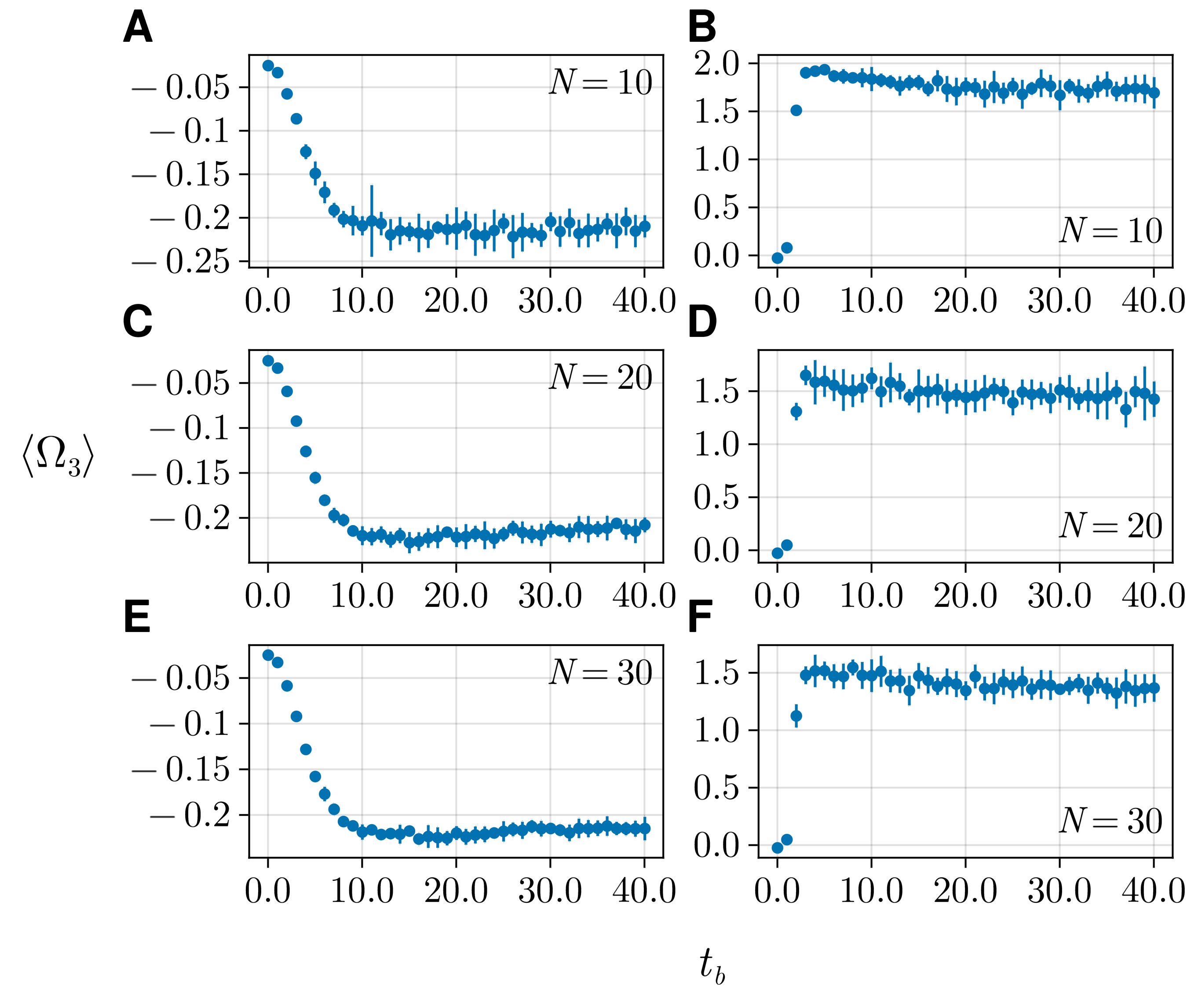}
    \caption[Sensitivity analysis for varying $\beta$]{
        Sensitivity analysis with respect to inverse temperature $\beta$ for systems with fixed $\alpha=0.4$.
        Left column (\textbf{A}, \textbf{C}, \textbf{E}) heterophilous systems.
        Right column (\textbf{B}, \textbf{D}, \textbf{F}) homophilous systems.
        Rows correspond to $N=10$ (\textbf{A}, \textbf{B}).
        Middle row correspond to $N=20$ (\textbf{C}, \textbf{D}).
        Bottom row correspond to $N=30$ (\textbf{E}, \textbf{F}).
        The results shown for $\alpha=0.4$ exemplify the behaviour observed for other values of $\alpha$.
        Each value is the mean across 5 replica ensembles each of size $R=10^5$, and error bars denote standard deviation.
    }
    \label{paper3:fig:sensitivity-beta}
\end{figure}

\clearpage
\subsection[Sensitivity against replica ensemble size]{Sensitivity against replica ensemble size $R$}
\begin{figure}[H]
    \centering
    \includegraphics[width=.70\linewidth]{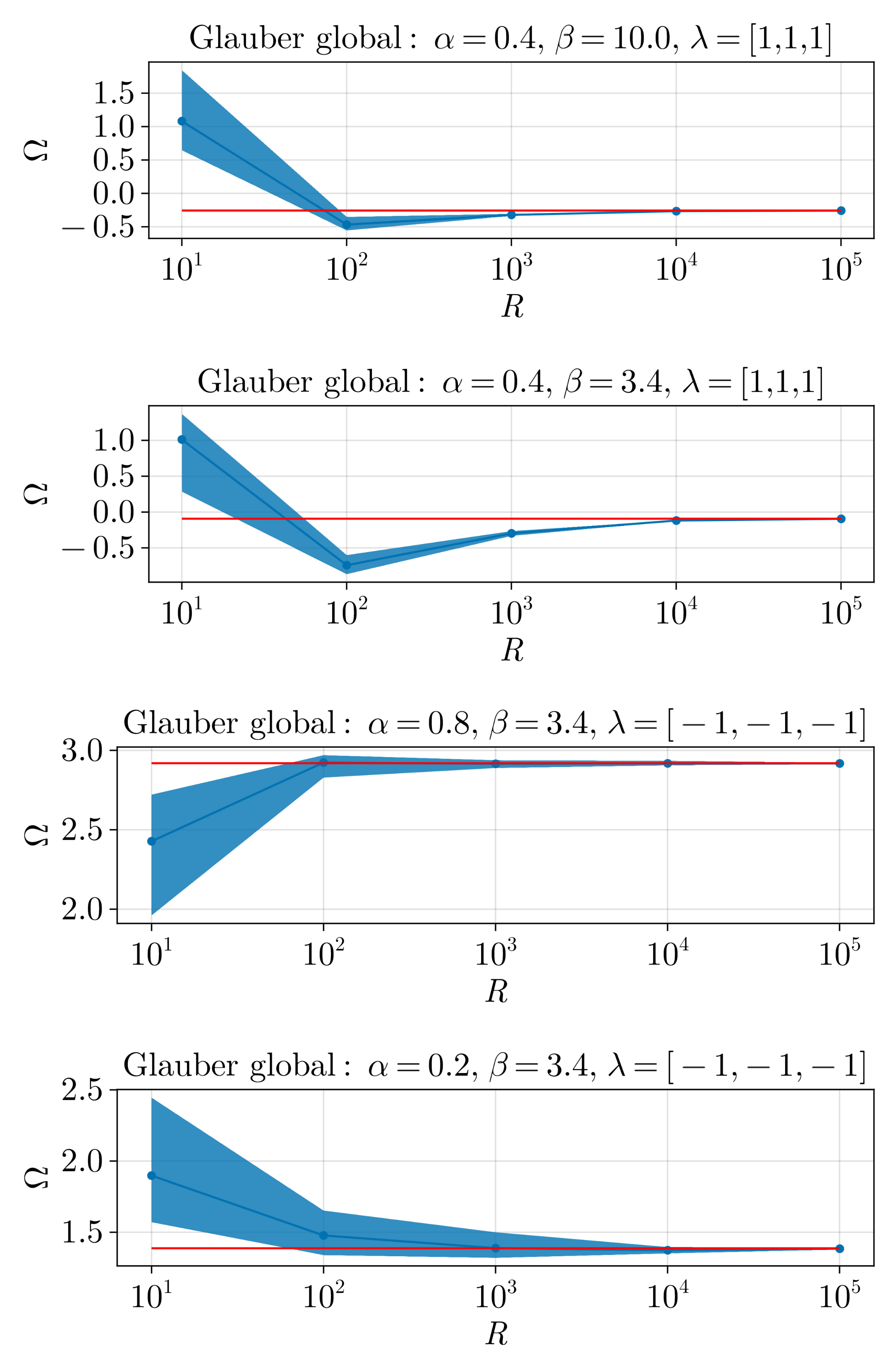}
    \caption[Sensitivity analysis for varying replica ensemble size]{
        \footnotesize Sensitivity analysis with respect to replica ensemble size $R$.
        Red lines corresponds to the exact value of $\Omega$ computed from the Boltzmann distribution with inverse temperature $\beta$. 
        For each replica we take a snapshot of the system state after $t_b=100$ sweeps.
    }
    \label{paper3:fig:sensitivity-ensemble-size}
\end{figure}
\clearpage
\section[Extended theoretical analysis of the N=3 model for arbitrary G]{Extended theoretical analysis of the $N=3$ model for arbitrary $G$}\label{paper3:sec:extended-IT-N3}
\subsection[Ground-state ensemble O-information for increasing G]{Ground-state ensemble O-information for increasing $G$}
\begin{figure}[!htbp]
    \centering
    \includegraphics[width=.99\linewidth]{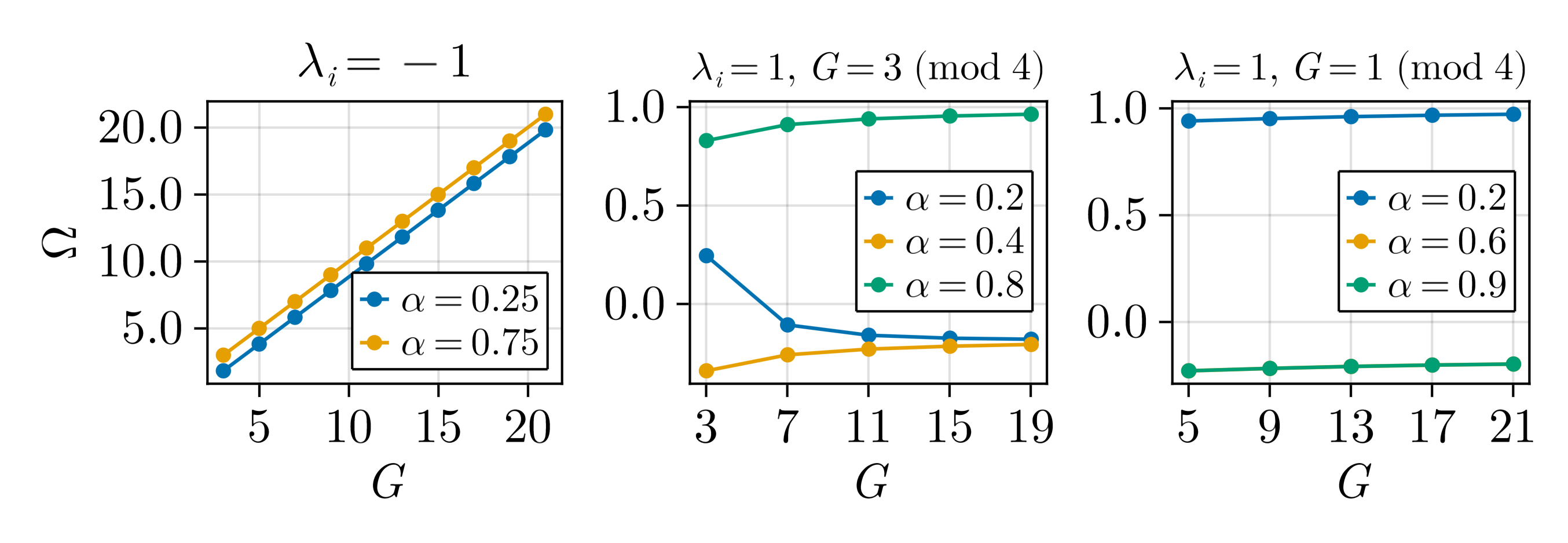}
    \caption[Ground-state ensemble O-information as a function of $G$ in systems of size $N=3$.]{
        Ground-state ensemble O-information as a function of $G$ in systems of size $N=3$.
        Note that, for $\lambda_i=1$ and $G=1$ ($\mathrm{mod}$ 4) values of $\Omega$ for $\alpha=0.6$ and $\alpha=0.9$ are equal.
    }
    \label{paper3:fig:ground-state-increase-G}
\end{figure}
In Fig.~\ref{paper3:fig:ground-state-increase-G} we show the O-information at zero-temperature (i.e., computed from the ground state ensemble) for increasing $G$.
For homophilous systems ($\lambda_i=-1$ for all $i$), the O-information increases monotonically as $G$ increases.
For heterophilous systems instead, the O-information quickly reaches a stable plateau.
Interestingly, for the heterophilous case with $G=3\;(\mathrm{mod}\;4)$, the ground state ensemble for $\alpha < 0.25$ is redundancy dominated only for $G=3$.
As $G = 3 \;(\text{mod }4)$ increases this becomes synergy-dominated.
\clearpage
\section{Case study extra figures}
\begin{figure}[H]
    \centering
    \includegraphics[width=.99\linewidth]{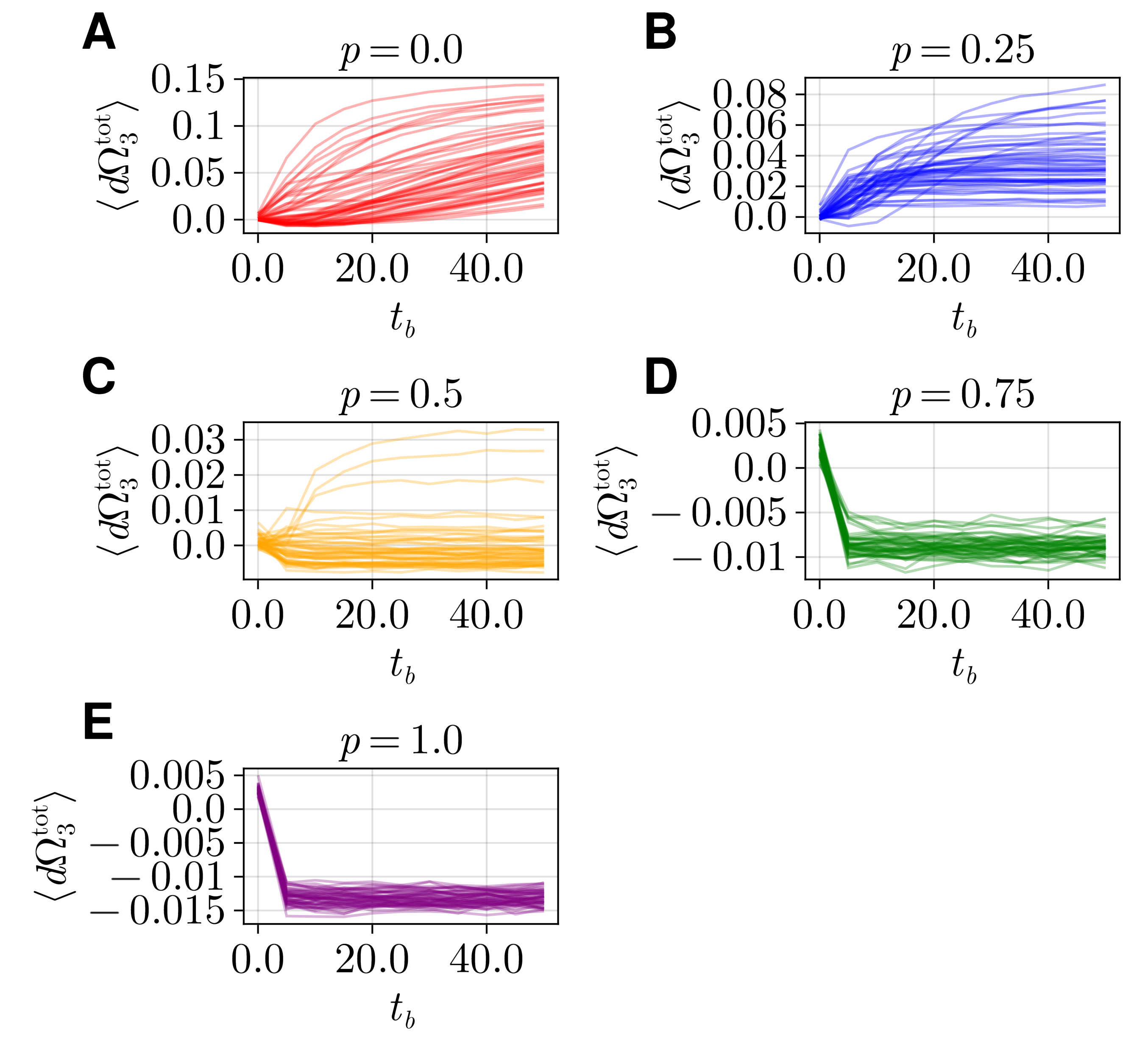}
    \caption[Individual trajectories of the total dynamical O-information]{
         Individual trajectories of the total dynamical O-information $\langle d\Omega^{\mathrm{tot}}_3\rangle$ as a function of $t_b$ for the case study of Fig.~\ref{paper3:fig:perturb-polarization}.
         Each trajectory corresponds to a distinct run.
         Each panel shows trajectories for different fractions of heterophilous agents $p$ after the perturbation: (\textbf{A}) $p=0$, (\textbf{B}) $p=0.25$, (\textbf{C}) $p=0.5$, (\textbf{D}) $p=0.75$, and (\textbf{E}) $p=1$.
         System parameters are as in Fig.~\ref{paper3:fig:perturb-polarization}.
    }
    \label{paper3:fig:polarization-extra}
\end{figure}

\end{document}